\documentclass[]{pasj02} 
\usepackage[switch,mathlines]{lineno} 
\usepackage{natbib} 
\usepackage[version=4]{mhchem}

\jyear{2026}

\newcommand{\beq}{\begin{equation}}
\newcommand{\eeq}{\end{equation}}
\newcommand{\beqn}{\begin{eqnarray}}
\newcommand{\eeqn}{\end{eqnarray}}

\begin{document}

\title{UV-driven organic precursor formation and dust evolution outside the snow line}

\author{
Ivan Judah \textsc{Manoraj}\altaffilmark{1}\altemailmark\email{manoraj.i.be10@m.isct.ac.jp},
Satoshi \textsc{Okuzumi}\altaffilmark{1}\orcid{0000-0002-1886-0880}, and 
Kazuaki \textsc{Homma}\altaffilmark{1}
}

\altaffiltext{1}{Department of Earth and Planetary Sciences, Institute of Science Tokyo, 2-12-1 Ookayama, Meguro, Tokyo 152-8551, Japan}


\KeyWords{planets and satellites: formation --- protoplanetary disks --- astrochemistry} 

\maketitle

\begin{abstract}
In the outer regions of protoplanetary disks, dust grains are often mantled with a layer of ice containing simple molecules. The irradiation of the disk by far-ultraviolet (FUV) photons breaks such molecules, producing photoproducts (radicals, ions, and intermediary molecules) in the ice layer that can act as precursors to complex organics. As grains drift towards the warmer, inner region of the disk, their consequent sublimation of ices causes organic synthesis on their surfaces, potentially enhancing grain sticking and dust growth. In addition to interstellar medium radiation, FUV emission from the central star and nearby massive stars can contribute to photoproduct formation. We perform a vertical dust size evolution simulation including grain growth and fragmentation, and calculate photoproduct formation under irradiation from these three sources, treating the central star's FUV continuum and Lyman-$\alpha$ emission separately. We find that enhanced UV irradiation, particularly from stellar Lyman-$\alpha$ emission, significantly increases the total abundance of photoproducts formed. This results from both the strong stellar Lyman-$\alpha$ flux and resonant scattering by the overlying hydrogen layer, which redistributes photons and directs a fraction toward the dust layer. Turbulent mixing and fragmentation cycles further redistribute grains vertically, allowing photoproducts to form throughout the disk. We find that stellar Lyman-$\alpha$ irradiation can produce sufficient photoproducts to produce an organic mantle with an {organic-to-rock ratio of $\approx 10\%$}, which could enhance grain sticking. Our results suggest that FUV-driven organic synthesis may facilitate dust growth and contribute to overcoming barriers to planetesimal formation.

\end{abstract}


\section{Introduction}\label{sec: intro}

Stars {form} within cool, dense molecular clouds, and often in clusters. Surrounding these young stars are protoplanetary disks, structures made from gas and dust orbiting the star. These disks inherit much of their mass, size, and chemical composition from their broader environment, and are the environment in which planets form. 

Planets are thought to form from micron-sized dust particles within the disk that collide at sufficiently low velocities to stick and grow into aggregates. At this stage, surface forces such as van der Waals forces play a role in their coagulation \citep{dominik_physics_1997}. Larger aggregates then form planetesimals, bodies on the order of kilometers, through further coagulation and gravitational/streaming instabilities \citep{youdin_streaming_2005, drazkowska_planetesimal_2017, franceschi_constraining_2023}. Planetesimals can then accrete dust from their surroundings and/or collide with other similar-sized bodies to eventually form full-sized planets {\citep{greenberg_planetestimals_1978, kokubo_formation_2000, ormel_effect_2010, lambrechts_rapid_2012}; see \cite{ormel_planet_2026} for review}. 

However, the process of planetesimal formation is not straightforward. The fragmentation barrier poses significant limits on dust growth, especially apparent in turbulent disks {\citep{blum_experimental_1993}}. Dust aggregates can fragment upon collision depending on their relative collision velocity {\citep{dominik_physics_1997, blum_experiments_2000}; see \cite{birnstiel_dust_2024} for review}. Disks with average turbulence ($\alpha \sim 10^{-3}$) can have collision velocities of dust particles reaching tens of $\rm m~s^{-1}$ \citep{ormel_closed-form_2007}, hindering dust growth. Growth is further limited by the radial drift barrier, in which gas drag causes dust grains to lose angular momentum and spiral into the inner disk before they can form planetesimals \citep{weidenschilling_aerodynamics_1977, brauer_coagulation_2008}, and the bouncing barrier, where low-velocity collisions result in bouncing instead of sticking \citep{zsom_outcome_2010, arakawa_size_2023, oshiro_bouncing_2025}.

To overcome {such growth barriers}, studies have proposed that organic mantles on dust grains enhance adhesion by increasing sticking efficiency \citep{homma_rocky_2019}. This idea is based on experiments by \cite{kudo_role_2002}, who showed that the viscoelastic properties of organic material at $\sim$250 K promote grain sticking. \cite{piani_evolution_2017} reached similar conclusions. Organic mantles are commonly observed on cometary interplanetary dust particles \citep{flynn_organic_2013}, and molecular dynamics simulations of glycolic acid nanoparticles by \cite{anders_bouncing_2021} support the experimental results of \cite{kudo_role_2002}. In molecular clouds, \cite{kimura_cohesion_2015} argued that cohesion between organic-coated silicate grains is the most plausible coagulation mechanism, although they noted that nanoparticles can initially grow without organic coatings. Likewise, \cite{kimura_morphological_2020} found that organic aggregates have significantly higher fragmentation velocities than silicate or icy aggregates. However, other studies report conflicting results. \cite{bischoff_sticky_2020} measured the tensile strengths of five organic materials and found variations spanning four orders of magnitude, with more complex organics exhibiting lower tensile strengths than simpler ones. Similarly, \cite{nagaashi_surface_2025} found that synthetic insoluble organic matter has a surface energy comparable to, or lower than, silica, suggesting that enhanced sticking depends not only on surface energy but also on viscosity and environmental conditions such as temperature. Despite these uncertainties, organic mantles remain of particular interest for overcoming barriers to dust growth. \cite{homma_rocky_2019} in particular found that organic-mantled grains with an organic mass fraction of $\approx 5\%\,\mathrm{wt}$ or greater may grow beyond the fragmentation barrier into planetesimals through collisions.

A natural follow-up question is how organic matter forms in protoplanetary disks. Laboratory studies have shown that UV irradiation of astrophysical ice analogs produces a wide range of complex organic molecules (COMs), including amino acids and nucleobases \citep[e.g.][]{hagen_interstellar_1979, schutte_laboratory_1992, oba_nucleobase_2019}. In these experiments, gas mixtures are deposited onto cold substrates and irradiated with UV photons, which photodissociate simple molecules within the ice into reactive radicals. Organic matter is formed when the ices are warmed to room temperature, and its composition is analyzed by techniques such as mass/IR spectrometry or liquid chromatography. A host of COMs were discovered to be produced, including those as complex as amino acids \citep{munoz_caro_amino_2002, nuevo_detailed_2008}. By emulating astrophysical conditions that mirror the environment in protoplanetary disks, these experiments imply that organic compounds should naturally form on dust grains during disk evolution. Furthermore, the production of COMs is only weakly dependent on factors such as ice composition, radiation source, and warming rate \citep{ciesla_organic_2012, tenelanda-osorio_effect_2022}. Instead, the essential requirements are the presence of C, H, O, and N, temperatures low enough for ice condensation, and radiation energetic enough to dissociate the initial molecules \citep{benest_couzinou_journey_2024}.

While organic matter is always formed if the above-mentioned conditions are satisfied, the most important factor for photo-production is photon dosage. Previous works simulating astrophysical conditions in the laboratory assumed the UV flux to originate entirely from the interstellar medium (ISM). This radiation is provided by OB type stars, and takes the value of 1 Habing unit ($1~ G_{0}$) in the Solar neighborhood, or 10$^{8}~\rm photons~cm^{-2}~s^{-1}$ \citep{habing_interstellar_1968}. 

However, two additional sources of radiation warrant consideration. First, dust particles in a protoplanetary disk are also irradiated by the central star of the disk. Unlike photons arriving from the ISM, which are assumed to strike the disk perpendicular to its surface, photons coming from the central star generally intercept the disk at a small grazing angle, meaning that they may not penetrate the disk deep enough to have a significant effect on photoproduct formation. However, the emission of the central star can be categorized into two groups, Ly-$\alpha$ (denoted Ly-$\alpha$), and the broader FUV continuum. Ly-$\alpha$ photons are particularly important because many photochemical processes, including photodissociation, are strongly wavelength dependent \citep{van_dishoeck_photoprocesses_2006}. Consequently, Ly-$\alpha$ photons can dissociate species such as \ce{HCN}, \ce{OH}, \ce{H2O}, and \ce{H2CO}, while the closely related ions or radicals \ce{CN}, \ce{OH+}, \ce{H2O+}, and \ce{HCO+} remain unaffected.

Ly-$\alpha$ photons also interact with the disk differently from the FUV continuum. Following \cite{bethell_propagation_2011}, the disk can be divided into two layers: an atomic hydrogen layer (H-layer), and the traditional dust-defined FUV continuum irradiation surface, below which most of the disk mass lies. Because Ly-$\alpha$ is a hydrogen resonance line, these photons undergo resonant scattering in the H-layer, which redirects part of the stellar Ly-$\alpha$ flux downward. The scattered photons emerge isotropically and, due to the flared geometry of the H-layer, often strike the lower disk layer at large angles, allowing them to penetrate deeper than FUV continuum photons, which pass through the H-layer largely unaffected until reaching the dust irradiation surface. Consequently, Ly-$\alpha$ radiation can irradiate a larger fraction of dust grains and constitutes a considerable part of the total radiation that the dust particles will receive \citep{bethell_propagation_2011}. The role of Ly-$\alpha$ has also been studied by \cite{vaikundaraman_refractory_2025} in the context of destruction of carbon in the inner disk, highlighting the need to consider the spectral composition of the stellar radiation field in studies of photochemistry and carbon depletion. 

A second important UV source is nearby massive stars. Since stars commonly form in clusters \citep{adams_birth_2010}, protoplanetary disks are often exposed to enhanced radiation fluxes coming from other massive stars in the same or nearby clusters. This enhancement over the baseline ISM FUV photon flux can reach up to $10^{5}~ G_0$ \citep{ciesla_organic_2012}. Such enhanced irradiation can substantially increase photoproduct formation and the subsequent synthesis of organic matter.

The purpose of this study is to quantify the abundance of photoproducts formed via FUV irradiation in a protoplanetary disk as the dust evolves over time. We define \textit{photoproducts} as the fraction of irradiated molecules in the ice mantle that ultimately form refractory organic matter following the radial drift and warming of the ices. We take into account not only radiation from the ISM, but also the FUV continuum and Ly-$\alpha$ emission of the disk's central star, as well as radiation from nearby massive stars. We assess whether the resulting photoproduct abundance is sufficient to produce an organic coating after inward drift by comparing it with the organic mass fraction of $\approx 5\%$ (corresponding to an organic-to-rock ratio of $\approx 5\%$) required for enhanced grain sticking, as proposed by \cite{homma_rocky_2019}. We find that Ly-$\alpha$ emission from the central star produces photoproducts at a higher abundance than FUV radiation from the ISM alone and is sufficient to produce an organic mantle with an organic-to-rock ratio of $\approx 5\%$, whereas ISM irradiation alone is not. These results highlight the important role of the central star in driving organic synthesis in protoplanetary disks.

The structure of this paper is as follows: section \ref{sec: model} details our dust growth and radiative transfer models. Our results are shown in section \ref{sec: results}, followed by a discussion in section \ref{sec: discussion}. We summarize our main findings in section \ref{sec: conclusions}. 

\section{Model} \label{sec: model}
\begin{figure*}[htbp]  
  \includegraphics[width=\textwidth, bb = 0 0 960 540, trim=10 350 10 200, clip]{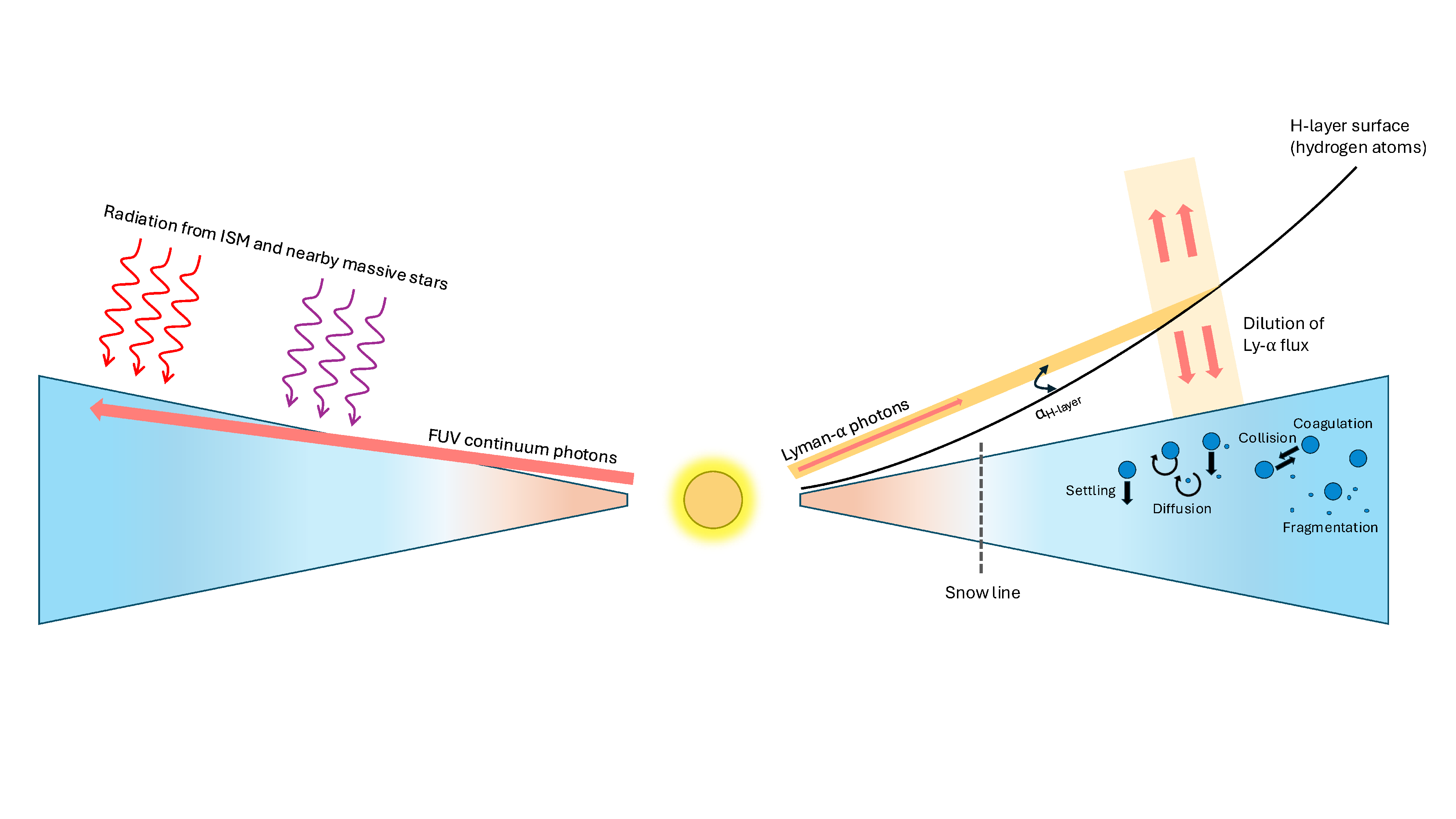}
  \caption{Schematic diagram of the model adopted in this study, showing the cross section of a disk and its central star. The illustration shows the various dust dynamics in our model, including settling, diffusion, collision, and fragmentation. The figure also shows the different sources of radiation onto the disk.}
  \label{fig disk processes}
\end{figure*}

Here we describe the model used in our simulations. We simulate the evolution of dust in a
protoplanetary disk and the creation of photoproducts via UV radiation. Specifically, we calculate the evolution of the vertical dust size distribution at a given radial distance in a disk taking into account coagulation, fragmentation, and vertical setting and diffusion (subsections~\ref{ssec: gas and dust model} and \ref{ssec: collisional growth and dust size evolution}). We simultaneously calculate the photon dosage to the grains from various types of UV radiation (subsections~\ref{ssec: Photoproduct formation} and \ref{ssec: UV radiation}). Figure~\ref{fig disk processes} presents a schematic of the adopted model.

\subsection{Gas and dust model}
\label{ssec: gas and dust model}

We adopt a minimum mass Solar Nebula model \citep{hayashi_structure_1981} around a solar-mass star. The radial gas surface density $\Sigma_{\rm g}$ is given as 
\begin{equation}
    \Sigma_{\rm g} = 19 \left( \frac{r}{20~\rm au} \right)^{-3/2}~\rm g~cm^{-2},
    \label{eq radial gas surface density}
\end{equation}
where $r$ is the distance from the central star, and the gas density $\rho_{\rm g}$ is given as
\begin{equation}
    \rho_{\rm g} = \frac{\Sigma_{\rm g}}{\sqrt{2\pi} h_{\rm g}} \exp \left[ -\frac{z^{2}}{2 h_{\rm g}^{2}} \right],
    \label{eq vertical density profile}
\end{equation}
where $h_{\rm g}$ is the gas scale-height and $z$ is the height from the disk mid-plane. The radial gas temperature profile around a solar-luminosity star is given by \cite{hayashi_structure_1981} as
\begin{equation}
    T = 62 \left( \frac{r}{20~\rm au} \right)^{-1/2} \, \rm K.
    \label{eq radial gas temperature profile}
\end{equation}
Our simulations are performed at a disk radius of 20 au. {We assume the gas disk is static and does not evolve with time.}

The dust particles initially present in the disk, dubbed monomers, are assumed to be composed of a silicate core with an ice mantle. We assume that the ice layer contains simple C-, H-, O-, and N-bearing species, which is one of the main requirements for organic formation \citep{ciesla_organic_2012}. Initially, the monomers are uniformly distributed in the disk with a constant dust to gas density ratio of $\rho_{\rm d}/\rho_{\rm g} = 0.01$.
The density of the silicate core is taken to be $\rho_{\rm si} = 2.6~\rm g~cm^{-3}$ \citep{chokshi_dust_1993}, and its radius is taken as $a_{\rm si} = 0.1~\rm \mu m$, the typical radius of chondritic porous interplanetary dust particles \citep{rietmeijer_size_1993}. The density of the ice mantle is simply assumed to be that of water, $\rho_{\rm ice} = 1~\rm g~cm^{-3}$. The total initial mass of the monomer $m_{0}$ is then given by
\begin{equation}
    m_{0} = \frac{4 \pi}{3} (a_{0}^{3} - a_{\rm si}^{3}) \rho_{\rm ice} + \frac{4 \pi}{3} a_{\rm si}^{3} \rho_{\rm si}
    \label{eq initial mass of monomer}
\end{equation}
where $a_{0}$ is the radius of the monomer (core + mantle). Assuming the ice-to-rock ratio of the monomer is unity \citep{lodders_solar_2003} yields $m_{\rm 0} \approx 2.18 \times 10^{-14}~\rm g$, and 
\begin{equation}
    a_{0} = a_{\rm si} \left( \frac{\rho_{\rm si}}{\rho_{\rm ice}} + 1 \right)^{1/3} \approx 0.153~\rm \mu m.
    \label{eq initial radius of monomer}
\end{equation}
The mean internal density of the monomers is
$\rho_{\rm 0} = 3m_0/(4\pi a_0^3) = 1.44~\rm g~cm^{-3}$.

\subsection{Collisional growth and dust size evolution}
\label{ssec: collisional growth and dust size evolution}

As monomers stick together, they grow into aggregates {\citep{dominik_physics_1997, guttler_outcome_2010}}. To examine the collisional growth of such monomers, we follow the model outlined in \cite{okuzumi_planetesimal_2012} and consider the case where a projectile aggregate of mass $m_{\rm p}$ collides with a target aggregate of mass $m_{\rm t}$, where $m_{\rm p} \leq m_{\rm t}$. Here, the aggregates are treated as compact spheres with radius $a$ and internal density $\rho_{\rm int} \approx \rho_{\rm 0}$. After the collision, the target's final mass is modeled as $m_{\rm t} + s(\Delta v, v_{\rm f})m_{\rm p}$, where $s(\Delta v, v_{\rm f})$, the dimensionless sticking efficiency, 
\begin{equation}
    s(\Delta v, v_{\rm f}) = \text{min} \left[ 1, -\frac{\ln(\Delta v/v_{\rm f})}{\ln5} \right],
    \label{eq sticking efficiency}
\end{equation}
is a function of the collision velocity $\Delta v$ and the fragmentation velocity $v_{\rm f}$. We refer to the newly formed aggregates as remnants, and the monomers as fragments. Aggregates undergo fragmentation depending on their collision velocity, with {fragmentation beginning at $\Delta v = v_{\rm f}/5$, where $s$ first drops below unity, and causing net mass loss of the remnants when $\Delta v > v_{\rm f}$, where $s$ becomes negative. Thus, $v_{\rm f}$ denotes the catastrophic fragmentation velocity \citep{blum_experiments_2000, guttler_outcome_2010, wada_collisional_2009, wada_growth_2013}.} The fragmentation velocity of the dust grains is taken as $v_{\rm f} = 1~\rm m\,s^{-1}$. This threshold is lower than the typical value of $\sim$ 10 $\rm m~s^{-1}$ for dust aggregates formed from water-ice mantled dust particles \citep{wada_growth_2013}. While early experiments suggested that icy dust grains are stickier than silicates and can form highly porous aggregates able to overcome the fragmentation barrier \citep{dominik_physics_1997, wada_collisional_2009}, this scenario has been challenged by more recent studies, which show that water ice is not sticky at low temperatures \citep{gundlach_tensile_2018, musiolik_contacts_2019}. The argument for low $v_{\rm f}$ is additionally supported by observations, such as those from \citet{okuzumi_nonsticky_2019} and \citet{ueda_support_2024}, who demonstrated that fragile dust better explains observations of disks around HL Tau and IM Lup. Therefore, we choose $v_{\rm f} = 1~\rm m~s^{-1}$ for our canonical case, and test $v_{\rm f} = 10~\rm m~s^{-1}$ in subsequent simulations. 

The velocity with which aggregates collide depends on the contribution of several other velocity factors: Brownian motion \citep{nakagawa_settling_1986}, radial and azimuthal drift, vertical settling \citep{adachi_gas_1976, weidenschilling_aerodynamics_1977}, and gas turbulence \citep{ormel_closed-form_2007}. Thus, we set the collision velocity as the square root of the sum of these contributions,
\begin{equation}
    \Delta v = \sqrt{\Delta v_{\rm B}^{2} + \Delta v_r^{2} + \Delta v_{\rm \phi}^{2} + \Delta v_z^{2} + \Delta v_{\rm tur}^{2}},
    \label{eq collision velocity}
\end{equation}
where $\Delta v_{\rm B}$ is the relative Brownian velocity, $\Delta v_r$ the relative radial drift velocity, $\Delta v_{\rm \phi}$ the relative azimuthal velocity, $\Delta v_z$ the relative vertical velocity, and $\Delta v_{\rm tur}$ the relative turbulence velocity, which are described in subsubsection 2.3.2 of \citet{okuzumi_rapid_2012}. {While we do not model the radial drift of dust grains (see below), we do consider the contribution of the relative radial drift velocity for the collision velocity.} These relative velocities (other than $\Delta v_{\rm B}$) are functions of the Stokes number, St, a dimensionless number that characterizes the behavior of particles in a fluid. It can be written as \citep{okuzumi_planetesimal_2012}
\begin{equation}
    \text{St} = \frac{\rho_{\rm int} a}{\rho_{\rm g}v_{\rm th}} \left(1 + \frac{4a}{9\lambda} \right) \Omega_{\rm K},
    \label{eq Stokes number}
\end{equation}
where $\rho_{\rm int}$ is the internal density of the dust aggregates, $a$ is the radius of the aggregates, $\lambda$ is the mean-free path of the gas molecules, and $v_{\rm th}$ is the thermal velocity.

We take into account the effects of coagulation, fragmentation, sedimentation, and vertical turbulent diffusion on the size and vertical distribution of dust particles, while neglecting the radial transport of the particles due to radial drift to focus on how the vertical dust evolution influences photoproduct yield. {The validity of neglecting radial drift is discussed in subsection \ref{ssec: implications for organic synthesis and dust coagulation}}. The dust size distribution function is given by $\mathcal{N}(z,m)$, defined as the dust number density per unit mass for aggregate mass $m$ at vertical height $z$ above the mid-plane. The evolution of $\mathcal{N}(z,m)$ is given by the vertical coagulation--advection--diffusion equation,
\begin{multline}
    \frac{\partial \mathcal{N}(z,m)}{\partial t} = -\frac{\partial}{\partial z} \left[ v_z\mathcal{N}(z,m) - D_{\rm d}\rho_{\rm g}\frac{\partial}{\partial z} \left( \frac{\mathcal{N}(z,m)}{\rho_{\rm g}} \right) \right] \\[0.5em] 
    + \frac{\partial \mathcal{N}(z,m)}{\partial t} \Bigg|_{\rm coll}.   
    \label{eq vertical coagulation advection diffusion equation for dust}
\end{multline}
Here, $v_z$ is the vertical settling velocity of the dust, given by 
\begin{equation}
    v_z = -\frac{\text{St}}{1 + \text{St}}z\Omega_{\rm K},
    \label{eq vertical settling velocity of dust}
\end{equation}
and $D_{\rm d}$ is the turbulence diffusion coefficient for dust aggregates \citep{youdin_particle_2007}, given by 
\begin{equation}
    D_{\rm d} = \frac{D_{\rm g}}{1 + \text{St}^{2}},
    \label{eq turbulence diffusion coefficient for dust}
\end{equation}
The turbulence diffusion coefficient for gas, $D_{\rm g}$, is given by 
\begin{equation}
   D_{\rm g} = \frac{\alpha c_{\rm s}^{2}}{\Omega_{\rm K}}.
   \label{eq diffusion coefficient in form of viscosity}
\end{equation}
where $\alpha$ is the Shakura--Sunyaev turbulence parameter \citep{shakura_black_1973}. We set $\alpha = 10^{-3}$ everywhere for our fiducial simulations, and also test $\alpha = 10^{-4}$ for certain runs. The first term on the right-hand-side of equation \eqref{eq vertical coagulation advection diffusion equation for dust} is the advection term, while the second is the diffusion term, and the last is the collision term, which includes the contribution of coagulation and fragmentation \citep{homma_isotopic_2024}. The collision term is given by
\begin{multline}
    m \frac{\partial \mathcal{N}(z,m)}{\partial t} \Bigg|_{\rm coll}   = -m \int_{0}^{\infty} dm'K(m,m')\mathcal{N}\mathcal{N'} \\[0.5em] +\int_{0}^{\infty}dm'\int_{0}^{m'}dm''K(m',m'')\mathcal{N'}\mathcal{N''} \\[0.5em] \times \left[ m\delta(m'+sm''-m) +(1-s)m''\delta(m_0 -m) \right],
    \label{eq collision term for dust}
\end{multline}
where $\mathcal{N}' = \mathcal{N}(z,m')$, and $\mathcal{N}'' = \mathcal{N}(z,m'')$. The collision rate coefficient, $K(z, m',m'')$, is given as 
\begin{equation}
    K(z,m',m'') = \sigma_{\rm coll} \Delta v,
    \label{eq vertically integrated collision rate coefficient}
\end{equation}
where $\sigma_{\rm coll} = \pi (a^2 + a'^2)$ is the collisional cross-section.

\subsection{Photoproduct formation}
\label{ssec: Photoproduct formation}

In laboratory experiments irradiating astrophysical ice analogs with ultraviolet radiation, COMs were formed when the ices were warmed to room temperature \citep[e.g.][]{munoz_caro_amino_2002, nuevo_detailed_2008}. As these experimental results have shown, warming the ice is a necessary step for organic synthesis \citep{bernstein_organic_1995}.

\begin{figure*}[htbp]
  {\centering
  \includegraphics[width=\textwidth, bb = 0 0 960 540, trim=10 50 10 70, clip]{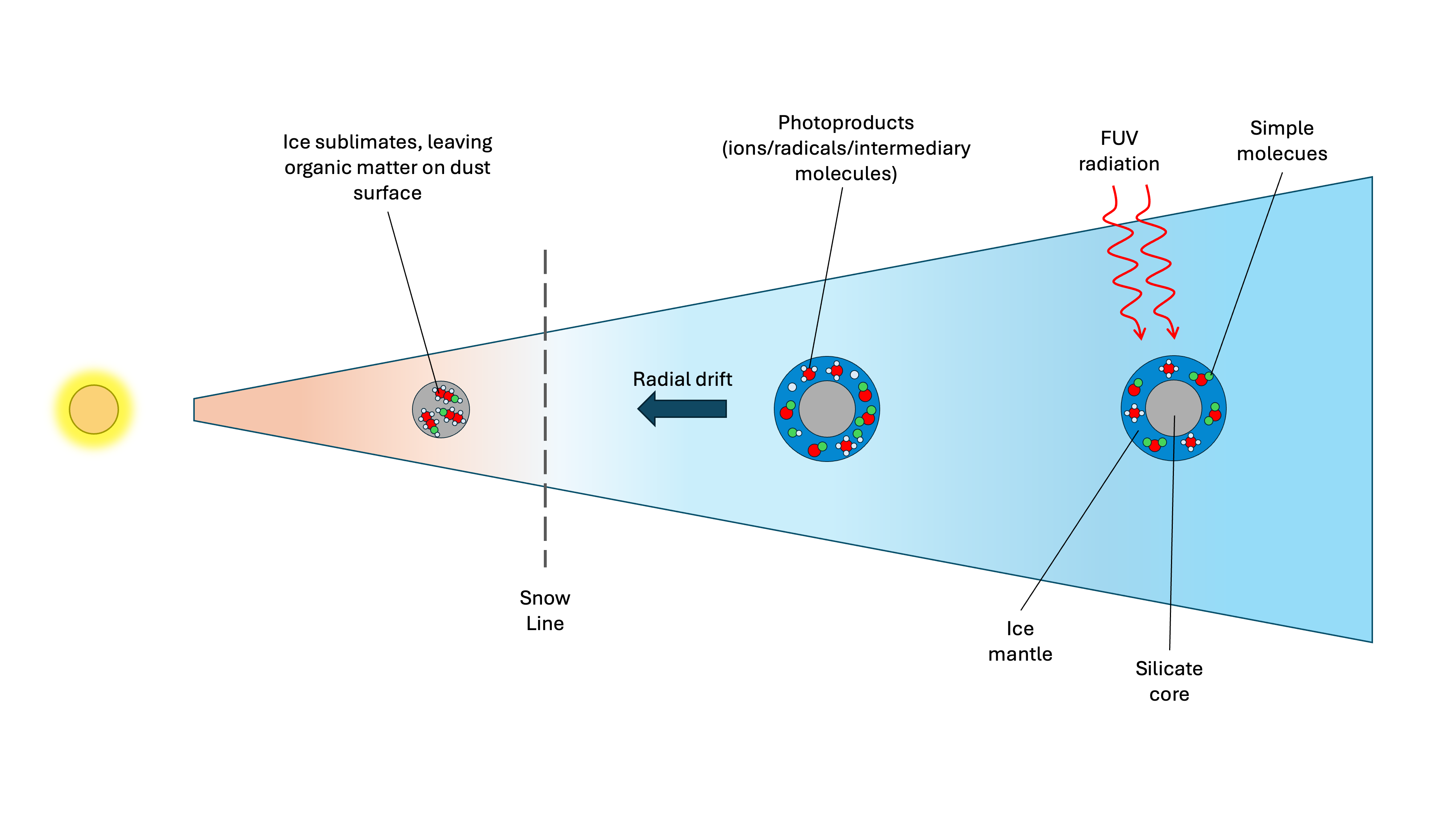}
  }
  \caption{Schematic diagram showing the process of organic matter formation from the FUV irradiation of icy dust grains. The diagram shows an ice mantled dust grain receiving FUV radiation. The radiation breaks down the simple molecules found in the mantle, producing photoproducts. As the grain drifts radially inward crossing the water ice snow line, the ice sublimates, triggering organic matter formation on the surface of the dust.}
  \label{fig dust radial drift}
\end{figure*}

In the context of a protoplanetary disk, this warming corresponds to the irradiated dust aggregates drifting into the warm, inner region of the disk. However, as mentioned in subsection \ref{ssec: collisional growth and dust size evolution}, we neglect this inward radial drift, and therefore calculate the abundance of \textit{photoproducts} present on the dust grains, before they would form organic matter \citep{homma_rocky_2019}. Not all molecules irradiated in the ice mantle will contribute to the formation of organic matter that remains on the dust grains, as some of the synthesized organics may be volatile and sublimate as the grains drift inward into warmer regions. Therefore, in this work, we define \textit{photoproducts} as the fraction of irradiated molecules that ultimately form refractory organic matter following the radial drift and warming of the ices. {Figure \ref{fig dust radial drift} shows a schematic of this process.}

To quantify {the photoproduct} abundance, we introduce the photoproduct mass density per unit aggregate mass, $\rho'_{\rm pp}$, as
\begin{equation}
    \rho'_{\rm pp}(z,m) = {c}(z,m)m\mathcal{N}(z,m),
    \label{eq photoproduct density}
\end{equation}
where ${c}(z,m)$ is the averaged photoproduct concentration. 

It is necessary to look at photoproduct formation in conjunction with grain size evolution as the processes that drive dust evolution also impact organic synthesis. For example, dust growth and fragmentation have consequences for the disk's overall opacity, which determines how deep UV photons are able to penetrate. Since photoproducts are formed on the surface of dust aggregates, we can expect $\rho'_{\rm pp}$ to evolve in tandem with the dust. 

We evolve $c(z, m)$ using the moment approach developed by \citet{okuzumi_numerical_2009} and later extended by \citet{homma_isotopic_2024} for dust evolution including fragmentation. In this approach, $c(z, m)$ obeys an evolutionary equation similar to equation \eqref{eq vertical coagulation advection diffusion equation for dust},
\begin{multline}
    \frac{\partial \rho'_{\rm pp}(z,m)}{\partial t} = 
    -\frac{\partial}{\partial z} \left[ v_z\rho_{\rm pp}'(z,m) 
    - D_{\rm d}\rho_{\rm g}\frac{\partial}{\partial z} 
    \left( \frac{\rho_{\rm pp}'(z,m)}{\rho_{\rm g}} \right) \right] \\[0.5em]
    + \frac{\partial \rho'_{\rm pp}(z,m)}{\partial t} \Bigg|_{\rm coll} 
    + \frac{\partial \rho'_{\rm pp}(z,m)}{\partial t} \Bigg|_{\rm sou}.   
    \label{eq vertical coagulation advection diffusion equation for photoproducts}
\end{multline}
The collision term in equation \eqref{eq vertical coagulation advection diffusion equation for photoproducts} is given by
\begin{multline}
    \frac{\partial \rho'_{\rm pp}(z,m)}{\partial t} \Bigg|_{\rm coll}   = -c(z,m) \int_{0}^{\infty} dm'K(m,m')\mathcal{N}\mathcal{N'} \\[0.5em] +\int_{0}^{\infty}dm'\int_{0}^{m'}dm''K(m',m'')\mathcal{N'}\mathcal{N''} \\[0.5em] 
    \times \left[ c_{\rm rem}\delta(m'+sm''-m) +c_{\rm frag}\delta(m_0 -m)\right],
    \label{eq collision term for photoproducts}
\end{multline}
where $c_{\rm rem}$ and $c_{\rm frag}$ are the average {concentrations} of photoproducts in the remnants and fragments. In reality, $c_{\rm rem}$ and $c_{\rm frag}$ depend on the mixing of monomers that occurs during the aggregate collision. However, as the details of this mixing process are not fully known, we follow \citet{homma_isotopic_2024} and simply assume that the mixing is efficient enough that the remnants and fragments have an equal photoproduct abundance, yielding the following relationship:
\begin{equation}
    \frac{c_{\rm rem}}{m'+sm''} = \frac{c_{\rm frag}}{(1-s)m''} \equiv F_{\rm new}.
    \label{eq equal photoproduct abundance}
\end{equation}
Here, $F_{\rm new}$ represents the photoproduct abundance of the remnants and fragments, and is determined by the conservation of the total number of photoproducts during the collision:
\begin{equation}
    F_{\rm new} = \frac{c(z,m')+c(z,m'')}{m' + m''}.
    \label{eq photoproduct abundance of remnants and fragments}
\end{equation}
Using this, equation \eqref{eq collision term for photoproducts} can be rewritten as
\begin{multline}
    \frac{\partial \rho'_{\rm pp}(z,m)}{\partial t} \Bigg|_{\rm coll}   = -c(z,m) \int_{0}^{\infty} dm'K(m,m')\mathcal{N}\mathcal{N'} \\[0.5em] +\int_{0}^{\infty}dm'\int_{0}^{m'}dm''K(m',m'')\mathcal{N'}\mathcal{N''} \\[0.5em] 
    \times F_{\rm new}\left[m\delta(m'+sm''-m) + (1-s)m'' \delta(m_0 -m)\right].
    \label{eq collision term for photoproducts with F_new}
\end{multline}

Equation \eqref{eq vertical coagulation advection diffusion equation for photoproducts} includes a source term to account for the formation of photoproducts via UV irradiation. {We do not explicitly model the chemical reactions occurring within the ice mantle during UV irradiation. Instead, w}e model the source term based on the work of \cite{jenniskens_carbon_1993}, who experimentally investigated the production efficiency of organic matter and found a linear relationship between the incident UV photon flux and the formation rate of organic molecules. The source term is written as 
\begin{equation}
    \frac{\partial \rho'_{\rm pp}(z,m)}{\partial t} \Bigg|_{\rm sou} = m_{\rm or} \epsilon F(z) \sigma_{\rm abs}(m) \mathcal{N}(z,m),
    \label{eq source term for organic matter}
\end{equation}
where $m_{\rm or}$ is the mass of an organic molecule, $\epsilon$ is the production efficiency (organic molecules per unit photon), $F(z)$ is the incident UV flux, and $\sigma_{\rm abs}$ is the absorption cross-section of the dust aggregate, approximated as its geometric cross-section, $\sigma_{\rm abs} = \pi a^{2}$. We approximate $m_{\rm or}$ as equal to the mass of a carbon atom, as carbon is the minimum component of organic matter. The production efficiency is given as $\epsilon = 4 \times 10^{-2}~\rm molecules~photon^{-1}$ by \cite{jenniskens_carbon_1993}. Similar experiments by \cite{bernstein_organic_1995} have yielded a production efficiency of $\epsilon = 2.5 \times 10^{-3}~\rm molecules~photon^{-1}$. We adopt the former parameter value.

Photo-production does not increase with radiation dose infinitely, but is ultimately limited by the exhaustion of the available starting materials in the ice. Following \cite{mumma_chemical_2011}, we assume that the composition of the ice mantle is similar to that of comets, and set the maximum photoproduct abundance as 10\% in weight of the mass of the ice. 

\subsection{UV radiation}
\label{ssec: UV radiation}
We consider four types of UV radiation in this work: the interstellar medium (ISM), the central star's FUV continuum and Ly-$\alpha$ emission, and nearby massive stars.

\subsubsection{ISM radiation}
\label{sssec: ism}

 The Solar System is exposed to an ionizing FUV flux that comes from O- and B-type stars in the ISM \citep{habing_interstellar_1968}. While the direction of this incoming radiation is set by the geometry between the disk and the source, for simplicity we can assume the flux reaches the disk perpendicular to its surface, allowing photons to penetrate into the disk and irradiate the dust within. However, the incident FUV flux $F(z)$ is attenuated as it travels from the surface of the disk to the mid-plane, as the photons are absorbed by the dust and gas in the disk. 

Let us consider an incident FUV flux hitting the disk at some angle $\theta$ to the disk surface. Ignoring gas absorption as well as scattering by gas and dust, this attenuation can be written as 
\begin{equation}
    \frac{dF}{ds} = - \int F \sigma_{\rm abs}(m) \mathcal{N}(z,m) dm,
    \label{eq fuv flux formula for ISM in terms of dS}
\end{equation}
where $F$ is the flux entering the disk, and $s$ is the path length along the line-of-sight of the ray. This equation can be rewritten in terms of the vertical direction to the disk surface, $dz = -ds \sin \theta$, as 
\begin{equation}
    \frac{dF}{dz} =  \frac{F}{\sin \theta} \int \sigma_{\rm abs}(m) \mathcal{N}(z,m) dm.
    \label{eq fuv flux formula for ISM in terms of dz}
\end{equation}
If we define the vertical optical depth from the disk surface $\tau_{\perp}(z)$ as
\begin{equation}
    \tau_{\perp}(z) = \int_{z}^{\infty} \left( \int \sigma_{\rm abs}(m) \mathcal{N}(z',m) \, dm \right) dz',
    \label{eq optical depth from disk surface}
\end{equation}
we can further rewrite equation \eqref{eq fuv flux formula for ISM in terms of dz} as
\begin{equation}
    \frac{dF}{d\tau_{\perp}} =  -\frac{F}{\sin \theta}. 
    \label{eq fuv flux formula for ISM in terms of dtau}
\end{equation}
The solution to equation \eqref{eq fuv flux formula for ISM in terms of dtau} is given by
\begin{equation}
   F(\tau_{\perp}) =  F(\tau_\perp = 0) \exp \left[ -\frac{\tau_{\perp}}{\sin \theta} \right].
    \label{eq solved fuv flux formula}
\end{equation}

For FUV radiation coming from the interstellar medium, the value of $F(\tau_\perp = 0)$ is set as $F_{0} = 1~G_{0}$, where $G_{0}$ is the Habing unit, equal to $10^{8} ~ \rm photons ~ cm^{-2} ~ s^{-1}$ \citep{habing_interstellar_1968}. For simplicity, we assume the ISM radiation hits the disk surface perpendicularly ($\theta = \pi/2$), and equation \eqref{eq solved fuv flux formula} takes the simple form
\begin{equation}
   F(\tau_{\perp}) =  F_{0} e^{-\tau_{\perp}}.
    \label{eq simple solved fuv flux formula for ISM}
\end{equation}
In reality, the ISM radiation field is approximately isotropic, and the approximate value of $\sin \theta$ is 1/2. Thus our perpendicularly-incident ISM approximation differs only by a factor of around two from the isotropic case and is adequate for an order-of-magnitude estimate.

\subsubsection{Central star continuum UV radiation}
\label{sssec: central star continuum}

In addition to the radiation coming from the ISM, a protoplanetary disk is also exposed to the flux coming from its own central star. The stellar FUV flux can be divided into two components: the broader FUV continuum and Lyman-alpha (Ly-$\alpha$) photons. A great part of the FUV spectrum of the central star is dominated by a strong, broad Ly-$\alpha$ line \citep{bethell_propagation_2011}. Ly-$\alpha$ photons, with wavelength $1215.67~ \text{\AA}$, are characteristically different from the broader FUV continuum. We follow \cite{bethell_propagation_2011} and assign a nominal FUV luminosity of $0.01~L_{\rm \odot}$ to the central star. Ly-$\alpha$ accounts for 80\% of the total FUV luminosity, while the remaining 20\% is FUV continuum photons. 

{The equation describing the FUV continuum flux in the disk from the central star is
\begin{equation}
   F(\tau_{\perp}) =  \frac{L_{\rm FUV, \, cont}}{4\pi r^{2}} \exp \left(-\frac{\tau_{\perp}}{\sin \theta} \right),
    \label{eq simple solved fuv flux formula for central star}
\end{equation}
where $L_{\rm FUV, \, cont}$ is the far-ultraviolet continuum luminosity of the star, and $\theta$ is the angle at which the radiation reaches the disk surface, called the grazing angle. The grazing angle is given by \cite{chiang_spectral_1997} as
\begin{equation}
     \theta = r \frac{d}{dr}\left ( \frac{z_{s}}{r} \right ),
     \label{eq grazing angle formula}
\end{equation}
where $z_{\rm s}$ is the height of the irradiation surface, estimated as $z_{s} = bh_{\rm g}$. With a parameter value of $b=3$, we calculate the grazing angle to be $\theta = 0.05$.} We calculate an incident flux of $4.2 \times 10^3 ~ G_{0}$ at a radius of 20 au, assuming a solar luminosity of $L_{\rm \odot} = 3.8 \times 10^{33} ~ \rm erg~ s^{-1}$.

\subsubsection{Central star Ly-$\alpha$ radiation}
\label{sssec: central star Ly-alpha}

Compared to FUV continuum photons, Ly-$\alpha$ photons have different interactions with certain chemical species. For example, \ce{HCN}, \ce{OH}, \ce{H2O}, and \ce{H2CO} can all be dissociated by Ly-$\alpha$ photons, while the closely related species \ce{CN}, \ce{OH+}, \ce{H2O+}, and \ce{HCO+} cannot {\citep{van_dishoeck_photoprocesses_2006}}. These wavelength-dependent chemical processes highlight their importance in determining what kinds of photoproducts can be produced {\citep{bergin_effects_2003}}. 

More importantly, Ly-$\alpha$ photons propagate toward the disk differently from FUV continuum photons \citep{bethell_propagation_2011}. Above the disk, a highly flared photodissociated hydrogen layer (H-layer) absorbs incident Ly-$\alpha$ photons and re-emits them isotropically, a process called resonant scattering. As a result, a fraction of the incident Ly-$\alpha$ flux is redirected downward toward the disk, where most of its mass resides. The scattered flux is diluted by a factor of $\sin\alpha_{\rm H\text{-}layer}$, where $\alpha_{\rm H\text{-}layer}$ is the angle at which the Ly-$\alpha$ photons are incident on the H-layer. Nevertheless, the downward-scattered Ly-$\alpha$ photons strike the disk at larger angles than FUV continuum photons, allowing them to penetrate deeper into the disk in the vertical direction. 

The fraction of Ly-$\alpha$ photons that are scattered downward depends on the scattering optical depth of the H-layer. Since the stellar radiation  grazes the disk at a small angle (i.e., $\alpha_{\rm H\text{-}layer} \ll 1$), the H-layer generally has a large line-of-sight (slant) optical depth  \citep{bethell_propagation_2011}. As a result, essentially all Ly-$\alpha$ photons incident on the layer undergo at least one resonant scattering event.
If the layer is vertically optically thin, approximately half of the singly scattered photons travel downward without undergoing further resonant scattering. If the layer is vertically optically thick, the photons undergo multiple scattering events, reducing the downward photon flux. 
In principle, the H-layer's optical thickness depends on the local abundance of small grains, with a lower dust abundance inhibiting \ce{H2} recombination, and thereby increasing the layer's thickness. However, \citet{bethell_propagation_2011} showed that this dependence is weak and that the layer's vertical optical thickness remains only modest ($\sim 6$) even when the abundance of small grains is depleted by a factor of 100 relative to the ISM value. We therefore assume that Ly-\(\alpha\) photons experience only a single scattering event in the H-layer.\footnote{Applying the two-stream approximation to radiative transfer in a purely scattering layer \citep{RybickiLightman79}, one can estimate the transmissivity of photons that are incident on one side of the layer at a small grazing angle to be ${\cal T} \approx 1/(2+\sqrt{3}\tau_{\rm layer})$, where $\tau_{\rm layer}$ is the layer's normal optical thickness. For $\tau_{\rm layer} \approx 6$, this gives ${\cal T} \approx 0.1$, which is a factor of five lower than the value ${\cal T} = 0.5$ for no multiple scattering. However, this order-of-unity uncertainty in the downward Ly-$\alpha$ flux has little effect on the rate of photoproduct formation because a significant fraction of the Ly-$\alpha$ photons in our computational domain is wasted on icy grains that are already saturated with photoproducts (see subsections~\ref{ssec: turbulence and fragmentation velocity dependence} and \ref{ssec: the role of photoproduct saturation}).}

Approximating the incident angles of the downward scattered Ly-$\alpha$ photons onto the disk as nearly $90^\circ$, their flux within the disk can be written as
\begin{equation}
    F(\tau_{\perp}) = \frac{1}{2} \left( \frac{L_{\rm Ly\alpha}}{4\pi r^{2}} \sin \alpha_{\rm H\text{-}layer} \right) e^{-\tau_{\perp}},
    \label{eq lyman alpha flux from central star}
\end{equation}
where $L_{\rm Ly\alpha}$ is the Ly-$\alpha$ luminosity of the central star. The factor of $1/2$ in equation \eqref{eq lyman alpha flux from central star} comes from the simple assumption that half the Ly-$\alpha$ photons that are processed by the H-layer are scattered downward.
We take $\sin \alpha_{\rm H\text{-}layer} \approx 1/20$, which is based on the ratio between the line-of-sight and vertical optical depths of the H-layer estimated by \citet{bethell_propagation_2011}. We calculate an incident flux of $4.2 \times 10^2 ~ G_{0}$ at a radius of 20 au. Although Ly-$\alpha$ emission constitutes a greater part of the total FUV luminosity than the FUV continuum, the incident Ly-$\alpha$ flux is lower due to the dilution factor upon striking the H-layer.

\subsubsection{Nearby massive stars}
\label{sssec: nearby massive stars}

Protoplanetary disks may also be exposed to radiation coming from nearby massive stars \citep{ciesla_organic_2012}, and this enhanced flux can reach up to $10^{3} \text{--} 10^{5}~G_{0}$ \citep{winter_external_2022}. Since most stars form in clusters \citep{adams_birth_2010}, disks around young stars are often illuminated by the other massive stars in the cluster. An example of this is the Trapezium Cluster. The most massive star in the cluster is the $\theta^{1}$ Orionis C star, which is itself a binary. As a highly luminous star, it emits a massive flux of ionizing UV radiation, enhancing the FUV flux received by protoplanetary disks in its vicinity. By approximating that this radiation are incident onto the disk nearly perpendicularly, we can write its flux within the disk as
\begin{equation}
   F(\tau_{\perp}) =  F_{\rm NMS} e^{-\tau_{\perp}},
    \label{eq simple solved fuv flux formula from nearby massive star}
\end{equation}
where $F_{\rm NMS}$ is the incident flux from nearby massive stars. We choose a representative value of $F_{\rm NMS}$ = $10^{4}~G_{0}$ for our simulation. 

\subsection{Simulation runs}
\label{ssec: simulation runs}

Our simulation runs are broadly divided into four categories based on the source of irradiation, and are listed in table \ref{parameter table}. We treat the ISM case as the baseline to compare our subsequent cases against. 

In order to investigate the effect of disk parameters, we varied the fragmentation velocity $v_{\rm f}$ and turbulence $\alpha$ according to the table and compared the results between the ISM and Ly-$\alpha$ cases. In addition, we conducted a handful of simulations removing the maximum limit on photoproduct formation to investigate the linear effect of increasing flux alone. These are referred to in the table as \texttt{no\_lim} simulations.

Each simulation was conducted at a distance of 20 au from the central star. This distance was chosen as it is outside the snow lines of \ce{CO2} and \ce{CH3OH}, where the required starting materials for organic formation have condensed into the icy mantle of dust grains \citep{sirono_sintering_2011}.

\begin{table}[!t]
 \begin{center}
 \caption{Adopted parameters for our simulation runs. The irradiation flux is the flux incident on the disk surface. \texttt{no\_lim} refers to simulations where the limit on the maximum amount of photoproducts that can be produced was removed.}
    \begin{tabular}{lcccc} 
      \hline
      Name & Incident Flux [$G_{0}$] & $v_{\rm f}~[ \rm m~s^{-1}]$ & $\alpha$ \\  
      \hline \\

      \texttt{ism\_fid}                     & 1                     & 1  & $10^{-3}$ \\
      \texttt{ism\_a4}                      & 1                     & 1  & $10^{-4}$ \\
      \texttt{ism\_vf10}                    & 1                     & 10 & $10^{-3}$ \\
      \texttt{ism\_no\_lim}                 & 1                     & 1  & $10^{-3}$ \\
      \texttt{ism\_a4\_no\_lim}             & 1                     & 1  & $10^{-4}$ \\
      \texttt{ism\_vf10\_no\_lim}           & 1                     & 10 & $10^{-3}$ \\
      \texttt{cs\_fuv\_fid}                 & $4.2 \times 10^3$     & 1  & $10^{-3}$ \\
      \texttt{cs\_fuv\_no\_lim}             & $4.2 \times 10^3$     & 1  & $10^{-3}$ \\
      \texttt{ly\_$\alpha$\_fid}            & $4.2 \times 10^2$     & 1  & $10^{-3}$ \\
      \texttt{ly\_$\alpha$\_a4}             & $4.2 \times 10^2$     & 1  & $10^{-4}$ \\
      \texttt{ly\_$\alpha$\_vf10}           & $4.2 \times 10^2$     & 10 & $10^{-3}$ \\
      \texttt{ly\_$\alpha$\_no\_lim}        & $4.2 \times 10^2$     & 1  & $10^{-3}$ \\
      \texttt{ly\_$\alpha$\_a4\_no\_lim}    & $4.2 \times 10^2$     & 1  & $10^{-4}$ \\
      \texttt{ly\_$\alpha$\_vf10\_no\_lim}  & $4.2 \times 10^2$     & 10 & $10^{-3}$ \\
      \texttt{nms\_fid}                     & $10^{4}$              & 1  & $10^{-3}$ \\
      \hline
    \end{tabular} 
    \label{parameter table}
 \end{center}
\end{table}

\section{Results} \label{sec: results}
In this section, we present the results of our simulations. We begin with the results of the fiducial models, comparing the organic yield for different radiation sources, before moving on to the results of simulations investigating turbulence and fragmentation velocity dependence. Finally, we present the results for simulations without the photoproduct limit.

\subsection{Fiducial models}
\label{ssec: fiducial models}

The results of our fiducial models (\texttt{ism\_fid}, \texttt{cs\_fuv\_fid}, \texttt{ly\_$\alpha$\_fid}, and \texttt{nms\_fid}) are consolidated into figure \ref{fig results ISM, CS FUV, Ly-a, NMS combined fiducial}. The first column plots the dust size distribution, which is the same for all models, while the subsequent columns plot the photoproduct distribution for each run. The dust size distribution is represented here as the dust aggregate density per decade of aggregate mass, $\Delta\rho_{\rm d}/\Delta \log m \equiv \ln(10)m^{2}\mathcal{N}(z,m)$. The photoproduct distribution is characterized by the local photoproduct-to-rock ratio $R(z,m) = 2c(z,m)$, where $c(z,m)$ is the averaged photoproduct concentration, defined in equation \eqref{eq photoproduct density}, and the factor of two comes from the assumption that rock composes half of the total grain mass.

We start with the first case of radiation coming from the interstellar medium (model \texttt{ism\_fid}). The first two columns of figure \ref{fig results ISM, CS FUV, Ly-a, NMS combined fiducial} show the results of this simulation. Initially, as the dust in the uppermost layers of the protoplanetary disk are irradiated, a burst of photo-production occurs, visible in the upper-left corners of the second column panels of figure \ref{fig results ISM, CS FUV, Ly-a, NMS combined fiducial}. As the dust evolves, coagulation of settled particles near the mid-plane leads to larger particles forming from the highly irradiated small particles, and the photoproduct distribution spreads downward. Mixing due to turbulence leads to a more homogeneous distribution closer to the mid-plane after $10^{6}$ yr. The final photoproduct distribution shows a portion of the dust (small particles near the surface of the disk) carrying a photoproduct-to-rock ratio of $\sim 10 \% ~ \rm wt$.

\begin{figure*}[htbp]
  {\centering
  \includegraphics[width=\textwidth]{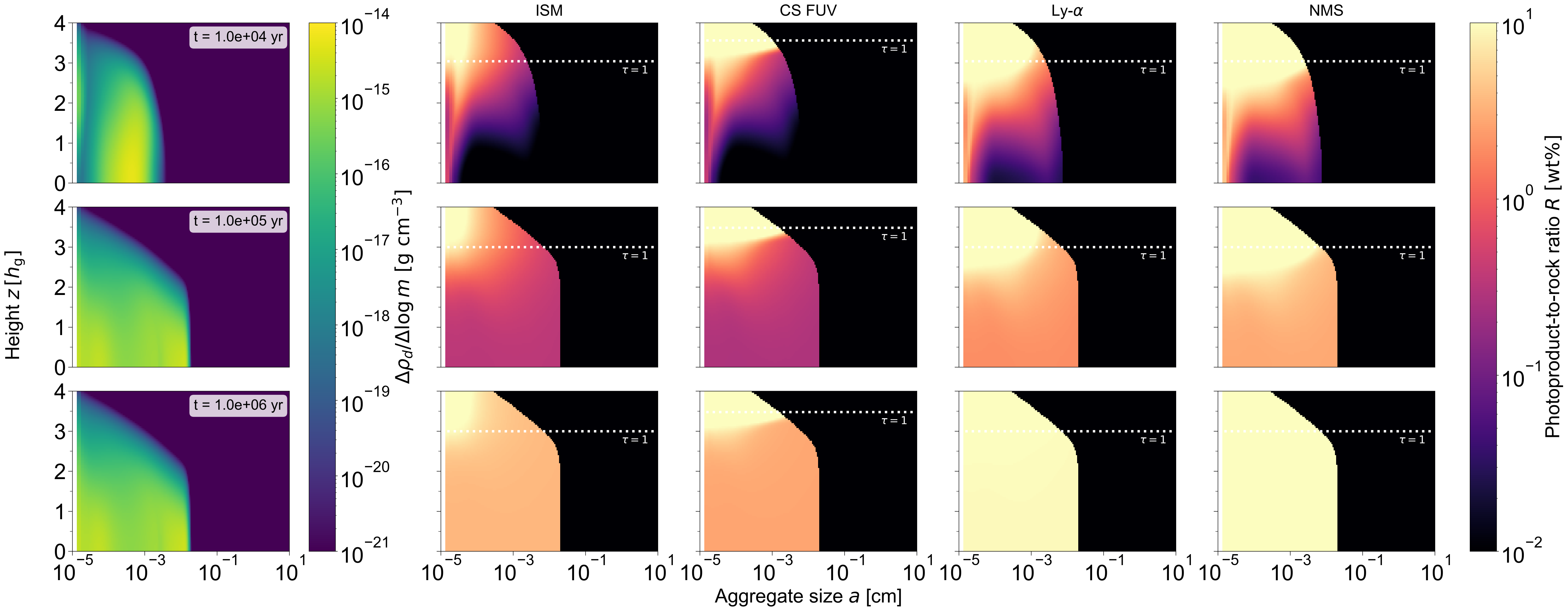}
  }
  \caption{Snapshots of the dust size distribution (first column) and photoproduct-to-rock ratio (second to fifth columns) for the \texttt{ism\_fid}, \texttt{cs\_fuv\_fid}, \texttt{ly\_$\alpha$\_fid}, and \texttt{nms\_fid} runs respectively, with all runs sharing the same dust size distribution. The top row shows the snapshot at $10^{4}$ yr, the middle at $10^{5}$ yr, and the bottom at $10^{6}$ yr. The y-axis shows the height above the mid-plane (normalized to the gas scale-height $h_{\rm g}$), and the x-axis plots the aggregate size. The dotted lines in columns two, four, and five show the location above the mid-plane where the vertical optical depth $\tau_{\perp} = 1$, whereas the dotted lines in column 3 show the location above the mid-plane where the line-of-sight optical depth $\tau_{\rm LOS} = 1$.}
  \label{fig results ISM, CS FUV, Ly-a, NMS combined fiducial}
\end{figure*}

The results of the central star continuum FUV emission run (model \texttt{cs\_fuv\_fid}) are shown by the first and third columns of figure \ref{fig results ISM, CS FUV, Ly-a, NMS combined fiducial}. As mentioned in subsubsection \ref{sssec: central star continuum}, FUV continuum photons experience absorption by the dust. Combined with their small grazing angle {of incidence} on the disk, they are unable to penetrate deep into the disk, leading to the stratified photoproduct distribution {between $3.0$ and $3.5~h_{\rm g}$} as seen in the panels of the third column of the figure. {To illustrate the penetration depth of the stellar FUV radiation, we also show the location where the line-of-sight optical depth $\tau_{\rm LOS}$ reaches unity. Here, $\tau_{\rm LOS}$ denotes the optical depth along the path of the incident stellar FUV continuum radiation and is related to the vertical optical depth by $\tau_{\rm LOS} = \tau_{\perp}/\sin\theta$, where $\theta$ is the angle of incidence measured relative to the disk surface. For our grazing angle, $\tau_{\rm LOS} \simeq 20\tau_{\perp}$.}

The first and fourth columns of figure \ref{fig results ISM, CS FUV, Ly-a, NMS combined fiducial} show the results for the Ly-$\alpha$ run (model \texttt{ly\_$\alpha$\_fid}). Unlike continuum FUV, Ly-$\alpha$ photons are able to penetrate through the depths of the disk due to their large grazing angle as a result of processing by the H-layer. This leads to higher abundances of photoproducts compared to the FUV continuum photon run, despite both runs having the same radiation source in the central star.

Finally, the results for the UV radiation coming from nearby massive stars run (model \texttt{nms\_fid}) are shown by the first and fifth columns of figure \ref{fig results ISM, CS FUV, Ly-a, NMS combined fiducial}. The photoproduct-to-rock distribution should follow that of the ISM run, as both types of radiation are treated the same, albeit with different magnitudes. Compared to the ISM run, the dust holds an overall higher abundance of photoproducts at any given time, simply due to the higher radiation dose provided by nearby massive stars. 

\begin{figure}
    {\centering
    \includegraphics[width=\linewidth, bb = 0 0 510.48 328.56]{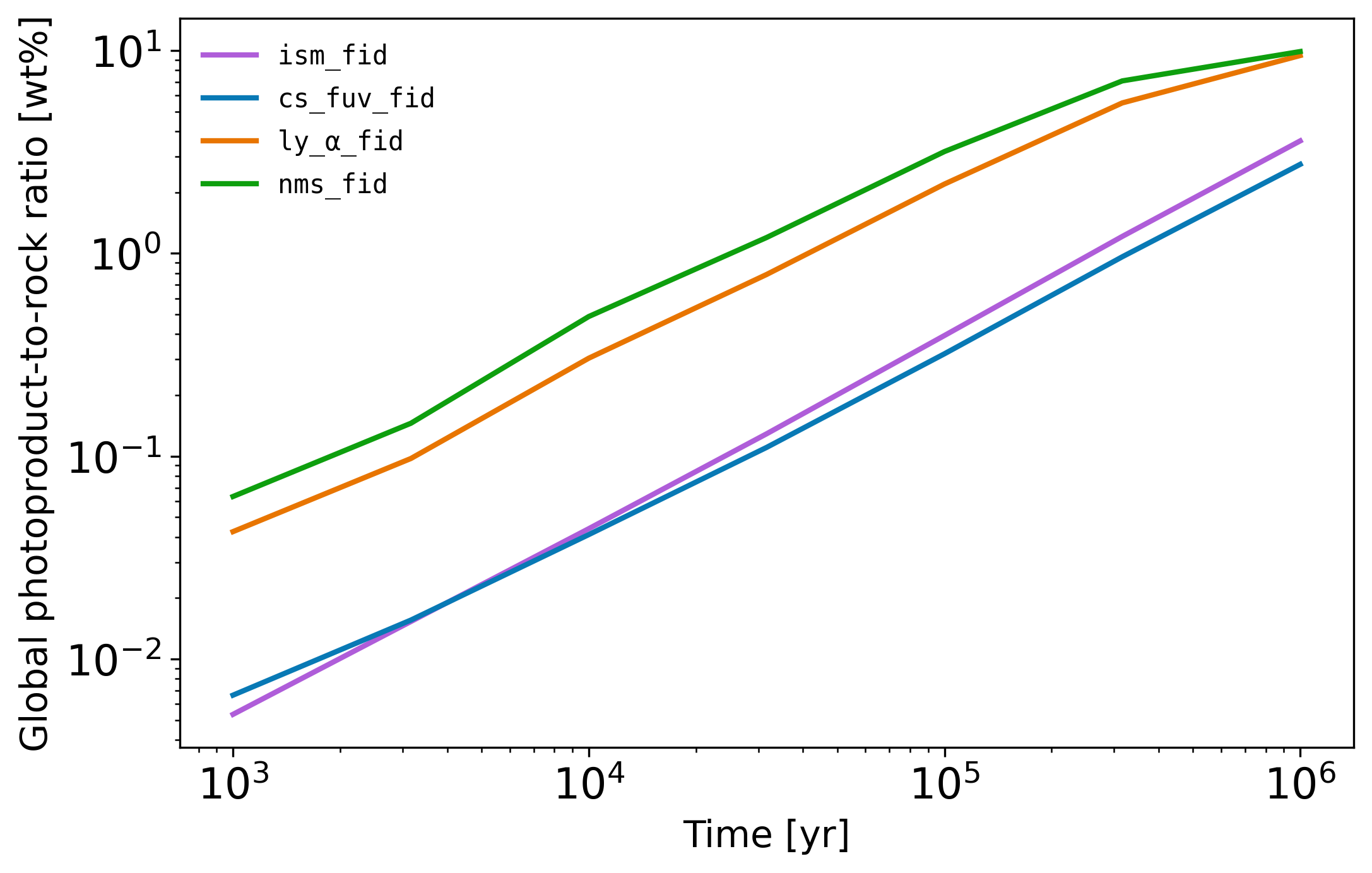}
    }
    \caption{Time evolution of the global photoproduct-to-rock ratio $R_{\rm global}$ for each source of radiation.}
    \label{fig global time evolution of photoproduct-to-rock ratio}
\end{figure}

Figure \ref{fig global time evolution of photoproduct-to-rock ratio} depicts the time evolution of the global photoproduct-to-rock ratio, defined as $R_{\rm global} = 2\Sigma_{\rm pp}/\Sigma_{\rm dust, tot}$, for the four models. Here, $\Sigma_{\rm pp} = \int (\int \rho'_{\rm pp} dm) dz$ and $\Sigma_{\rm dust, tot} = \int_{0}^{\infty}(\int m \mathcal{N}(z,m) dm )dz$ are the photoproduct surface density and total dust surface density, respectively. The factor of two converts the dust surface density into the rock surface density. The \texttt{nms\_fid} model produces the highest global photoproduct abundance, followed by the \texttt{ly\_$\alpha$\_fid} model. Both these models see their global photoproduct-to-rock ratios converge by $10^{6}$ yr, approaching the maximum value of $10\%$, which is the enforced maximum amount of photoproducts that can be formed in our model, based on the availability of the initial molecules in the ice mantle (see subsection \ref{ssec: Photoproduct formation}). The \texttt{ism\_fid} model yields a global photoproduct-to-rock ratio of $3.6\%$, short of the required value of $5\%$ for enhanced grain sticking, while the \texttt{cs\_fuv\_fid} model results in the lowest global photoproduct-to-rock ratio with {a value of $2.7\%$.}

While the flux coming from nearby massive stars is $10^4$ times higher than in the ISM model, their global photoproduct-to-rock ratios are separated by less than an order of magnitude. This is again due to the enforced limit on the maximum amount of photoproducts. The effects of the imposed photoproduct limit are discussed further in subsection~\ref{ssec: the role of photoproduct saturation}.

The difference of an order of magnitude between the global photoproduct-to-rock ratios of the Ly-$\alpha$ and central star FUV continuum runs is striking, given that both have the central star as their radiation source. This highlights the importance of the processing of Ly-$\alpha$ photons by the H-layer through resonant scattering, as mentioned in subsubsection \ref{sssec: central star Ly-alpha}. 

\subsection{Roles of turbulent diffusion and collisional fragmentation}
\label{ssec: turbulence and fragmentation velocity dependence}

The simulation results presented in figure \ref{fig results ISM, CS FUV, Ly-a, NMS combined fiducial} demonstrate that dust aggregates at the mid-plane are able to be irradiated through a cycle driven by fragmentation and turbulent diffusion. The initially unirradiated aggregates near the mid-plane release dust particles via fragmentation, which are transported to the disk surface by turbulent diffusion. At the disk surface, these particles are exposed to UV radiation, and are transported back downward as vertical turbulence mixes dust particles at the mid-plane with those near the surface. At the disk mid-plane, the irradiated dust particles grow or are incorporated into larger aggregates, before once again releasing more particles due to fragmentation, which subsequently experience the same cycle of diffusion and growth. This cycle allows the formation of highly irradiated macroscopic dust aggregates at the mid-plane, shown in the bottom panels of the \texttt{ism\_fid}, \texttt{ly\_$\alpha$\_fid}, and \texttt{nms\_fid} columns of figure \ref{fig results ISM, CS FUV, Ly-a, NMS combined fiducial}. 

Here, we invoke the diffusion timescale to check whether diffusion can be effective at $t > 10^{5}$ yr for the fiducial models. We calculate the diffusion timescale from the mid-plane to $\tau_{\perp} = 1$ as the dust particles are irradiated at $\tau_{\perp} = 1$. The diffusion timescale $t_{\rm diff}$ at $z$ can be estimated as 
\begin{equation} 
 t_{\rm diff}(z) = \frac{z^2}{D_{\rm d}} = \frac{(1+ {\rm St}^2)z^2}{\alpha c_{\rm s} h_{\rm g}} .
 \label{eq diffusion timescale}
\end{equation}
Taking the height at $\tau = 1$ as $3h_{\rm g}$ from our results, the diffusion timescale is given by $t_{\rm diff, \tau_{\perp}=1} = 9(1+ {\rm St}^2)/(\alpha \Omega_{\rm K})$. In our simulation, Stokes number is much lower than unity due to the low fragmentation threshold. Thus, the diffusion timescale is approximately given by $t_{\rm diff, \tau_{\perp}=1} \approx 9/(\alpha \Omega_{\rm K})$, which amounts to $\approx 10^5$ yr for the fiducial models with $\alpha = 10^{-3}$. This timescale is consistent with our results for models \texttt{ism\_fid}, \texttt{ly\_$\alpha$\_fid}, and \texttt{nms\_fid} shown in figure \ref{fig results ISM, CS FUV, Ly-a, NMS combined fiducial}, suggesting that dust at the mid-plane experiences irradiation as a result of turbulent diffusion in these runs.

In contrast to the above-mentioned three models, the \texttt{cs\_fuv\_fid} column of figure \ref{fig results ISM, CS FUV, Ly-a, NMS combined fiducial} shows that there remain poorly-irradiated aggregates near the mid-plane of the disk, despite the mixing effects of turbulence. This is because only a small amount of small dust grains lying at very high altitudes receive the entirety of the central star's FUV continuum radiation. This causes the ice on these grains to be fully saturated with photoproducts, namely, their photoproduct-to-rock ratios have already reached the upper limit of 10\%. Much of the incident FUV radiation is being wasted on these saturated dust grains, limiting the efficiency of photoproduct formation in this run.

\begin{figure}[t]
  {\centering
  \includegraphics[width=\linewidth, bb = 0 0 2307.6 1833.12]{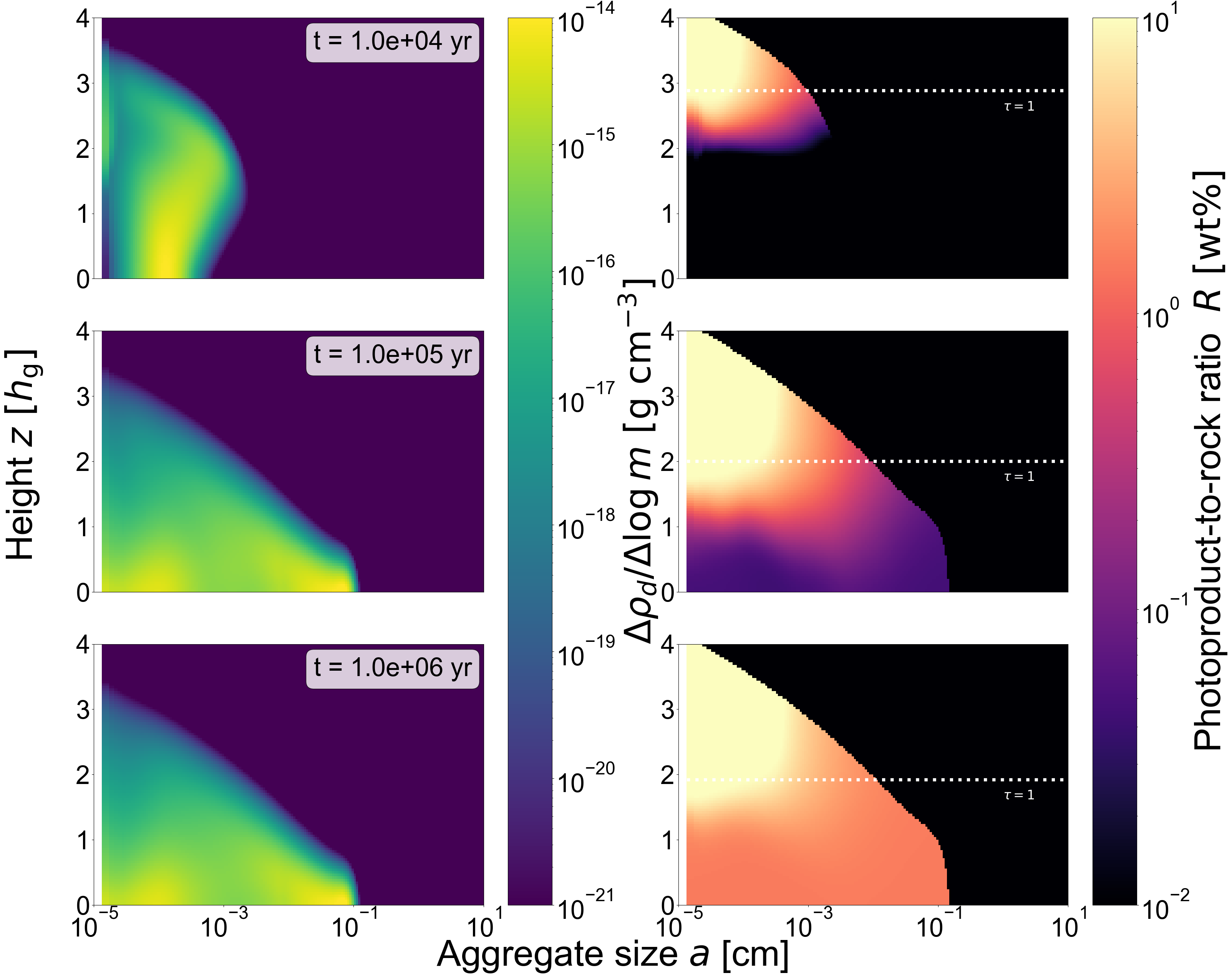}
  }
  \caption{Snapshots of the dust size distribution (left column) and photoproduct-to-rock ratio (right column) for the \texttt{ism\_a4} runs. The top row shows the snapshot at $10^{4}$ yr, the middle at $10^{5}$ yr, and the bottom at $10^{6}$ yr. The y-axis shows the height above the mid-plane (normalized to the gas scale-height $h_{\rm g}$), and the x-axis plots the aggregate size. The dotted line in the right column shows the location above the mid-plane where the vertical optical depth $\tau_{\perp} = 1$.}  
  \label{fig results ISM alpha4}
\end{figure}

\begin{figure}[t]
  {\centering
  \includegraphics[width=\linewidth, bb = 0 0 2290.32 1833.12]{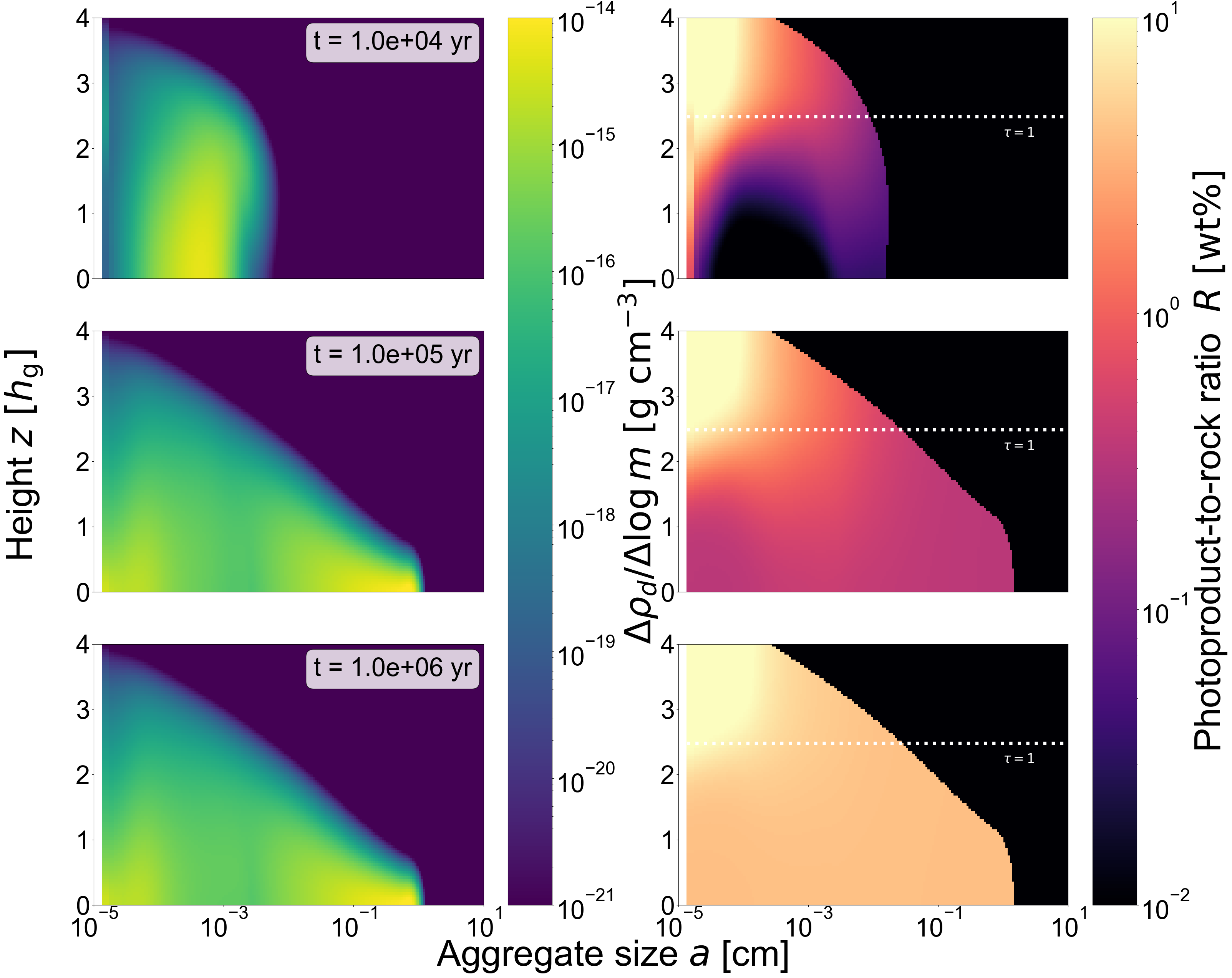}
  }
  \caption{Same as figure \ref{fig results ISM alpha4} but for the \texttt{ism\_vf10} run.}
  \label{fig results ISM vfrag10}
\end{figure}

In order to investigate the extent to which turbulence and fragmentation affect our results, we varied the Shakura-Sunyaev $\alpha$ prescription for turbulence and the fragmentation velocity $v_{\rm f}$ parameters according to table \ref{parameter table}, for the ISM and Ly-$\alpha$ models. The results for the ISM model are shown in figures \ref{fig results ISM alpha4} and \ref{fig results ISM vfrag10}. The biggest difference compared to the fiducial simulations can be seen in the dust size distribution (left panels of each figure). In the \texttt{ism\_fid} run, the maximum dust size after $10^{6}$ yr is 0.02 cm (first column of figure \ref{fig results ISM, CS FUV, Ly-a, NMS combined fiducial}), whereas in the \texttt{ism\_a4} run (figure \ref{fig results ISM alpha4}) and the \texttt{ism\_vf10} run (figure \ref{fig results ISM vfrag10}), the dust aggregates reach a maximum size of 0.15 cm and 1.5 cm respectively. The same is true of the corresponding Ly-$\alpha$ runs, as shown in figures~\ref{fig results Ly-a alpha4} and \ref{fig results Ly-a vfrag10} in appendix~\ref{sec:appendix}.

\begin{figure*}[t]
    \begin{center} 
        \includegraphics[width=0.45\linewidth, bb = 0 0 510.48 328.56]{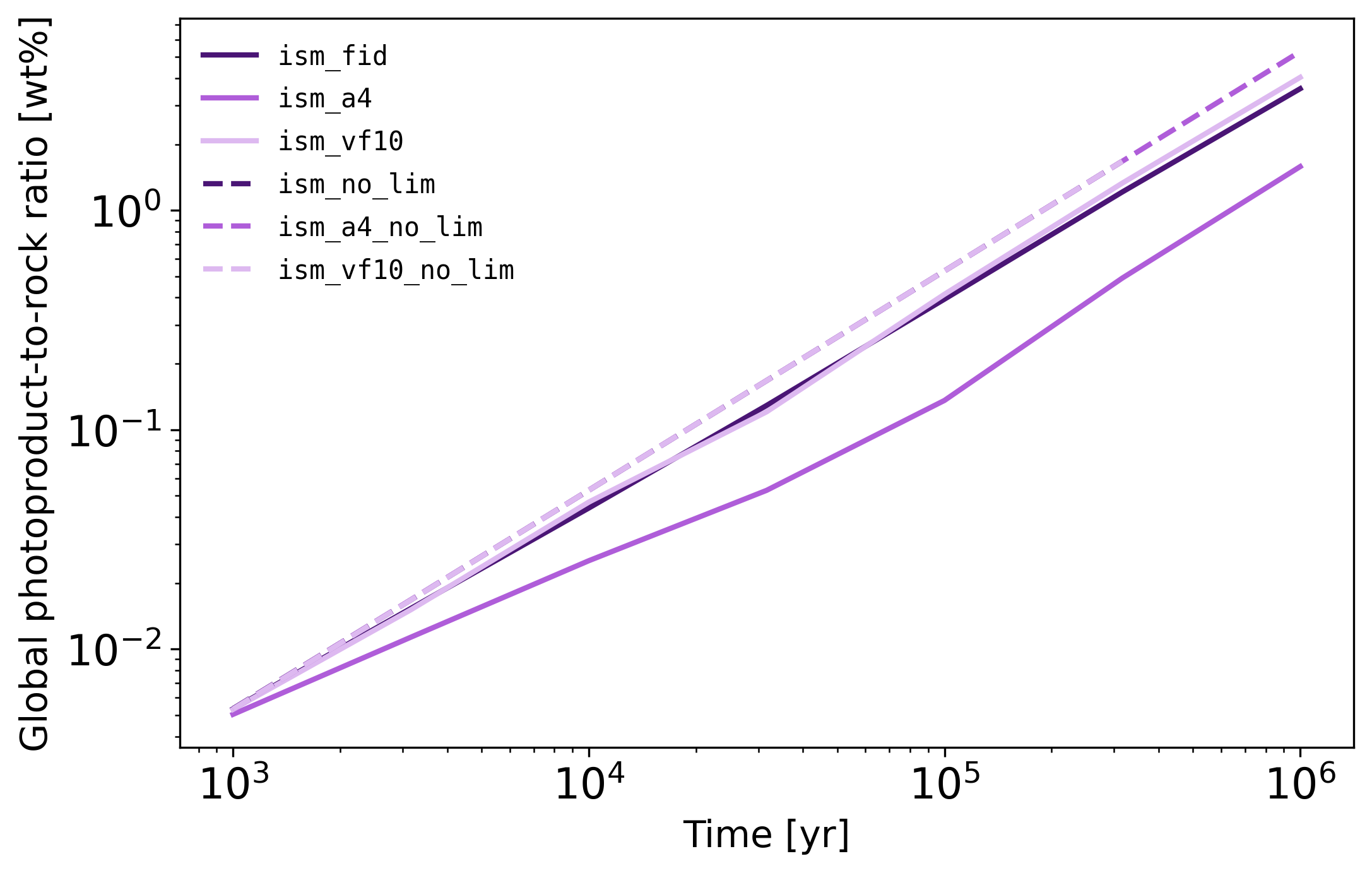}
        \hspace{2mm}
        \includegraphics[width=0.45\linewidth, bb = 0 0 510.48 328.56]{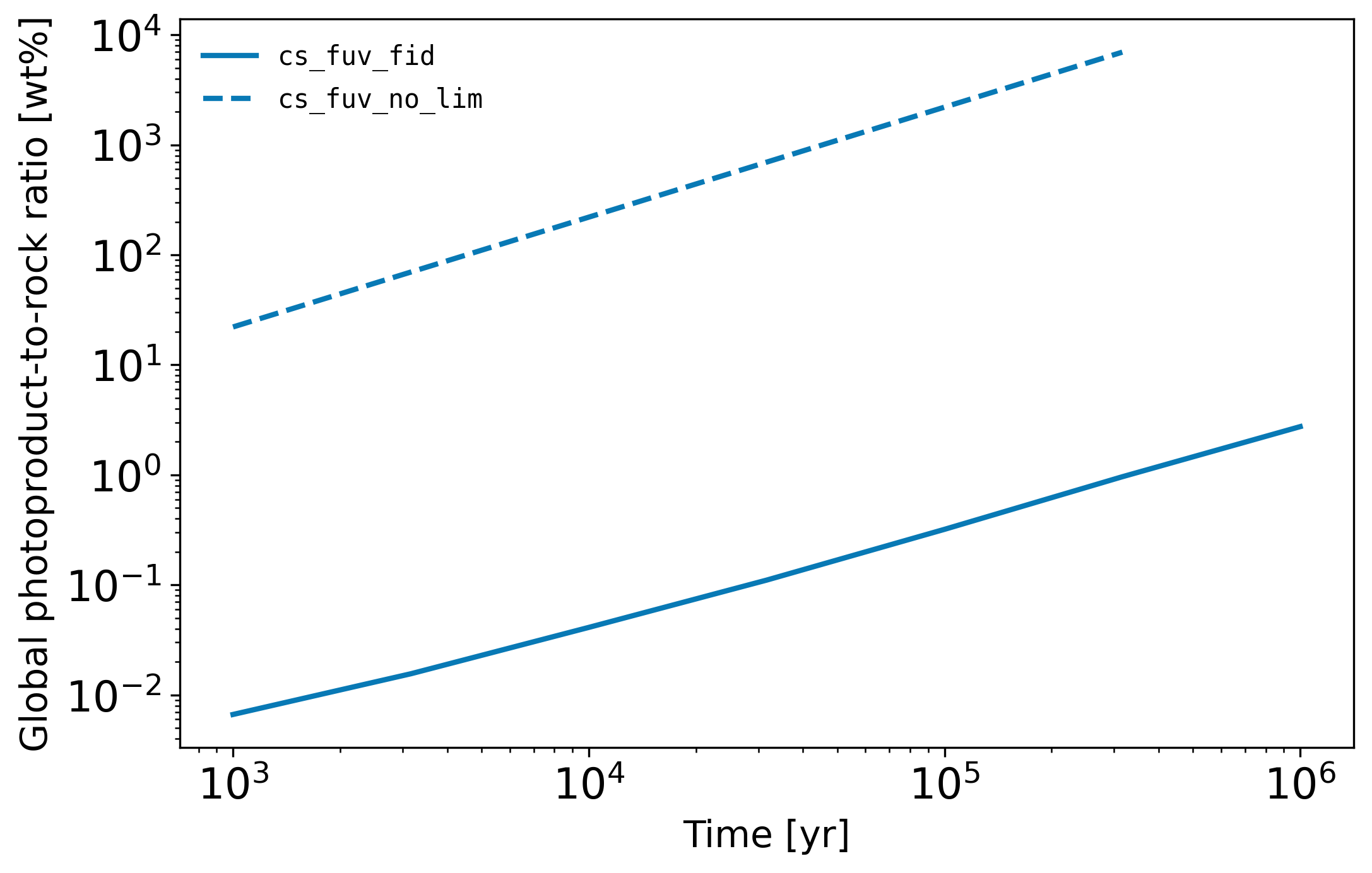}
    \vspace{1em}
        \includegraphics[width=0.45\linewidth, bb = 0 0 510.48 328.56]{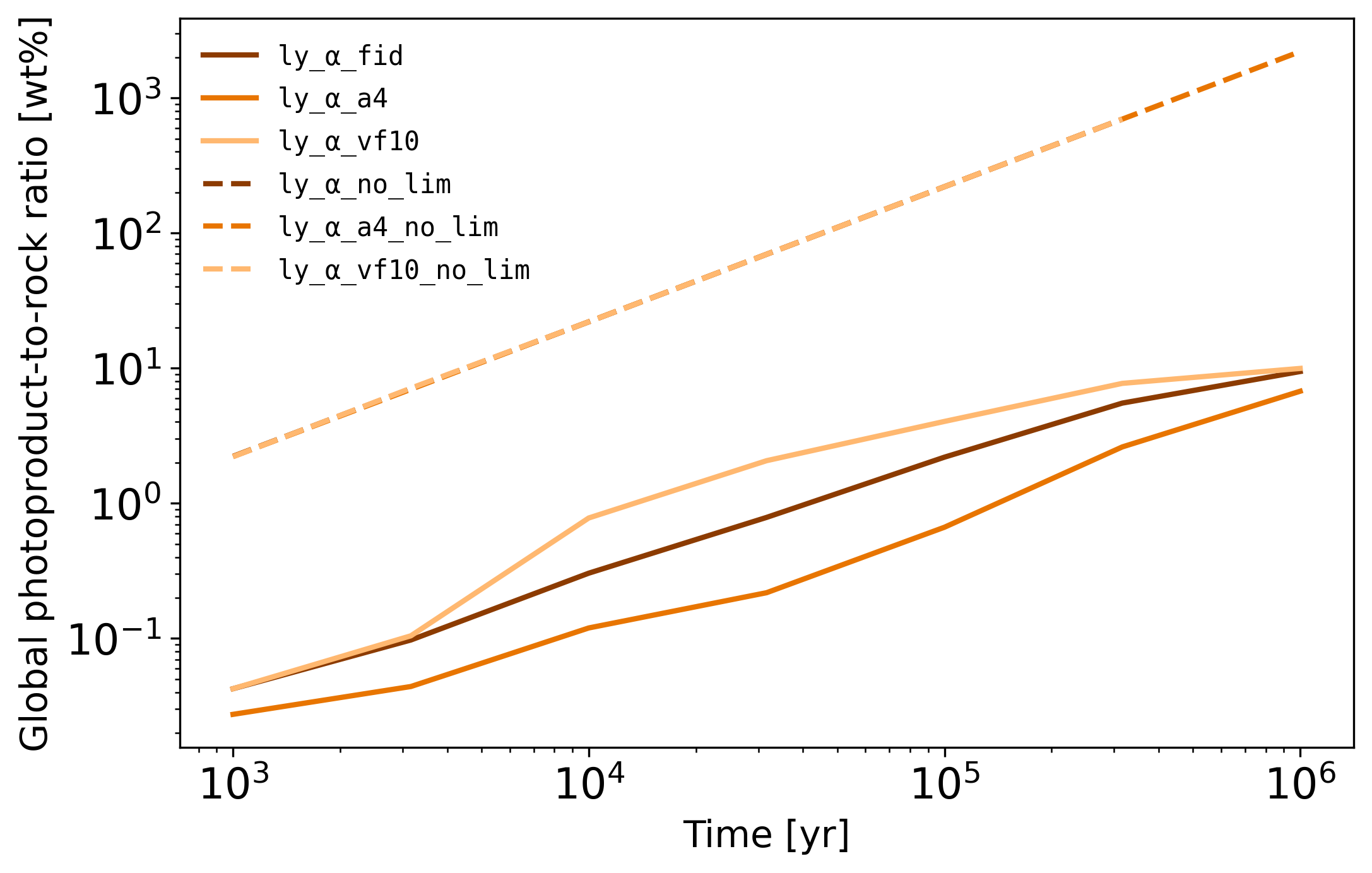}
        \hspace{2mm}
        \includegraphics[width=0.45\linewidth, bb = 0 0 510.48 328.56]{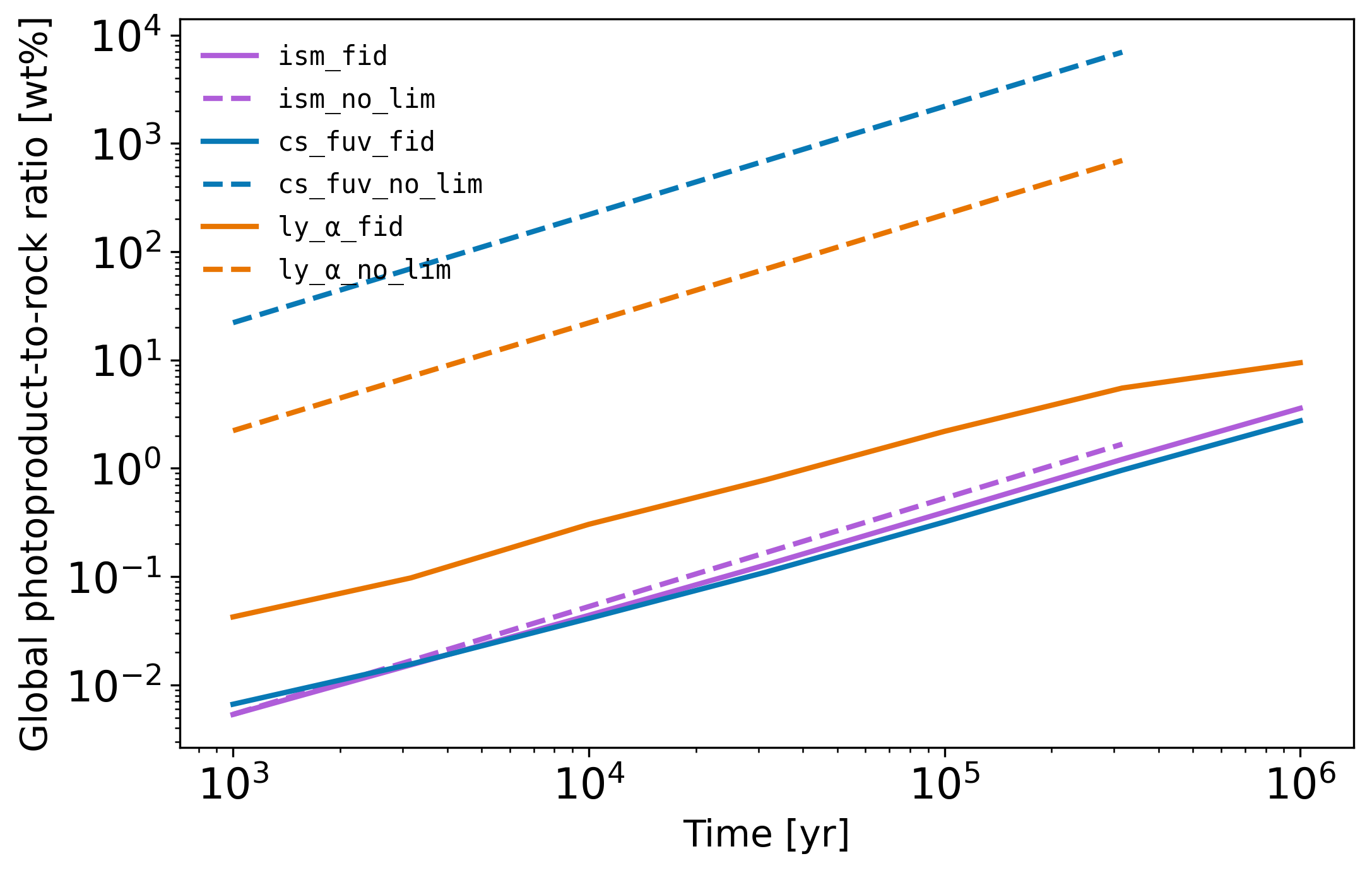}
    \end{center}
    \caption{Time evolution of the global photoproduct-to-rock ratio $R_{\rm global}$ for different radiation source simulations. Top left: ISM source; top right: central star FUV source; bottom left: Ly-$\alpha$ source; bottom right: comparison of all three. The dashed lines show the results of the \texttt{no\_lim} simulations.}
    \label{fig photoproduct-to-rock ratio time evolution all}
\end{figure*}

However, varying the turbulence and fragmentation velocity resulted in a difference of less than an order of magnitude in the final global photoproduct-to-rock ratio of the dust compared to the fiducial model. The solid lines in the upper and lower left panels of figure \ref{fig photoproduct-to-rock ratio time evolution all} depict the time evolution of the global photoproduct-to-rock ratio for ISM and Ly-$\alpha$ runs with different values of $\alpha$ and $v_{\rm f}$. At the end of $10^{6}$ yr, the global photoproduct-to-rock ratio for the \texttt{ism\_fid}, \texttt{ism\_a4}, \texttt{ism\_vf10} models are 3.6\%, 1.6\%, and 4.0\% respectively, all within the same order of magnitude. The global photoproduct-to-rock ratios for the \texttt{ly\_$\alpha$\_fid}, \texttt{ly\_$\alpha$\_a4}, and \texttt{ly\_$\alpha$\_vf10} models nearly converge to the imposed maximum value of 10\% by $10^{6}$ yr. This indicates that the global photoproduct-to-rock ratios in the ISM source models would also converge, albeit over a timescale longer than $10^{6}$ yr.

\subsection{The role of photoproduct saturation}
\label{ssec: the role of photoproduct saturation}

The maximum amount of photoproducts that can be produced via UV irradiation ultimately depends on the availability of starting molecules in the ice mantle of dust grains. This limit on the maximum amount of photoproducts is enforced in our simulations as 10\% in weight of the mass of the {ice}. However, in order to illuminate why the \texttt{cs\_fuv\_fid} run (third column of figure \ref{fig results ISM, CS FUV, Ly-a, NMS combined fiducial}) produces a lower global abundance of photoproducts than the \texttt{ly\_$\alpha$\_fid} model (fourth column of figure \ref{fig results ISM, CS FUV, Ly-a, NMS combined fiducial}) despite having a radiation flux an order of magnitude greater, we run simulations removing the limit on photoproduct formation. We caution the reader that these simulations are unrealistic as the photoproduct abundance in dust is allowed to exceed 100\%. 

The results of the \texttt{ism\_no\_lim}, \texttt{cs\_fuv\_no\_lim}, and \texttt{ly\_$\alpha$\_no\_lim} simulations are shown in figure \ref{fig results ISM, CS FUV, Ly-a combined no limit}, with additional results shown in figure \ref{fig results ISM, Ly-a combined alpha4 no limit} and \ref{fig results ISM, Ly-a combined vfrag10 no limit} in appendix~\ref{sec:appendix}. {Some simulations were terminated before reaching the nominal end time due to the lengthy computational time required. However, as shown in figure \ref{fig photoproduct-to-rock ratio time evolution all}, the evolution of the global photoproduct-to-rock ratio is effectively identical across the \texttt{no\_lim} runs within each radiation source group, such that the different final simulation times do not affect the comparison presented here.} In these simulations, the optically thick disk in our model absorbs all incident photons, resulting in photoproducts formed at a rate proportional to the irradiation flux alone. Turbulence and fragmentation velocity parameters have no bearing on the abundance of photoproducts formed. 

The dashed lines in the first three panels of figure \ref{fig photoproduct-to-rock ratio time evolution all} plot the time evolution of the global photoproduct-to-rock ratio for all the \texttt{no\_lim} runs, referenced in table \ref{parameter table}. As seen in the figures, the dashed lines in each graph overlap perfectly, indicating that photoproduct formation depends solely on the incident flux if the photoproduct limit were absent. The Ly-$\alpha$ models, with an incident flux of $420~ G_0$ (compared to the ISM flux of $1~ G_0$), result in a final global photoproduct-to-rock ratio 420 times larger than the ISM run. Similarly, the central star FUV models (top right panel) produce the highest global photoproduct-to-rock ratio compared to the previous two runs, due to its incident flux of $4200~ G_0$. 

The three radiation sources are compared in the bottom right panel of figure \ref{fig photoproduct-to-rock ratio time evolution all}. Note how the central star FUV runs (blue lines) are spaced the farthest apart. Despite having the largest flux out of the three models, as long as the photoproduct limit is in place, FUV continuum radiation coming from the central star will produce the lowest abundance of photoproducts due to the small grazing angle of the incident radiation and the inefficient downward transport of photons within the disk. Ly-$\alpha$ and ISM radiation, with their lower incident flux but perpendicular incidence, produce a higher global photoproduct-to-rock ratio (solid lines in the figure). However, once the photoproduct limit is removed, and photoproduct generation depends solely on flux, the central star FUV continuum radiation produces the highest global photoproduct-to-rock ratio (dashed lines).

These simulation runs illustrate the importance of enforcing a photoproduct limit, as it is a physical effect present in real disks, determined by the finite amount of ice carried by individual dust grains and the molecules within the ice. The global increase in photoproducts is ultimately governed by the limited starting material available. 

\begin{figure*}[htbp]
  {\centering
  \includegraphics[width=\textwidth]{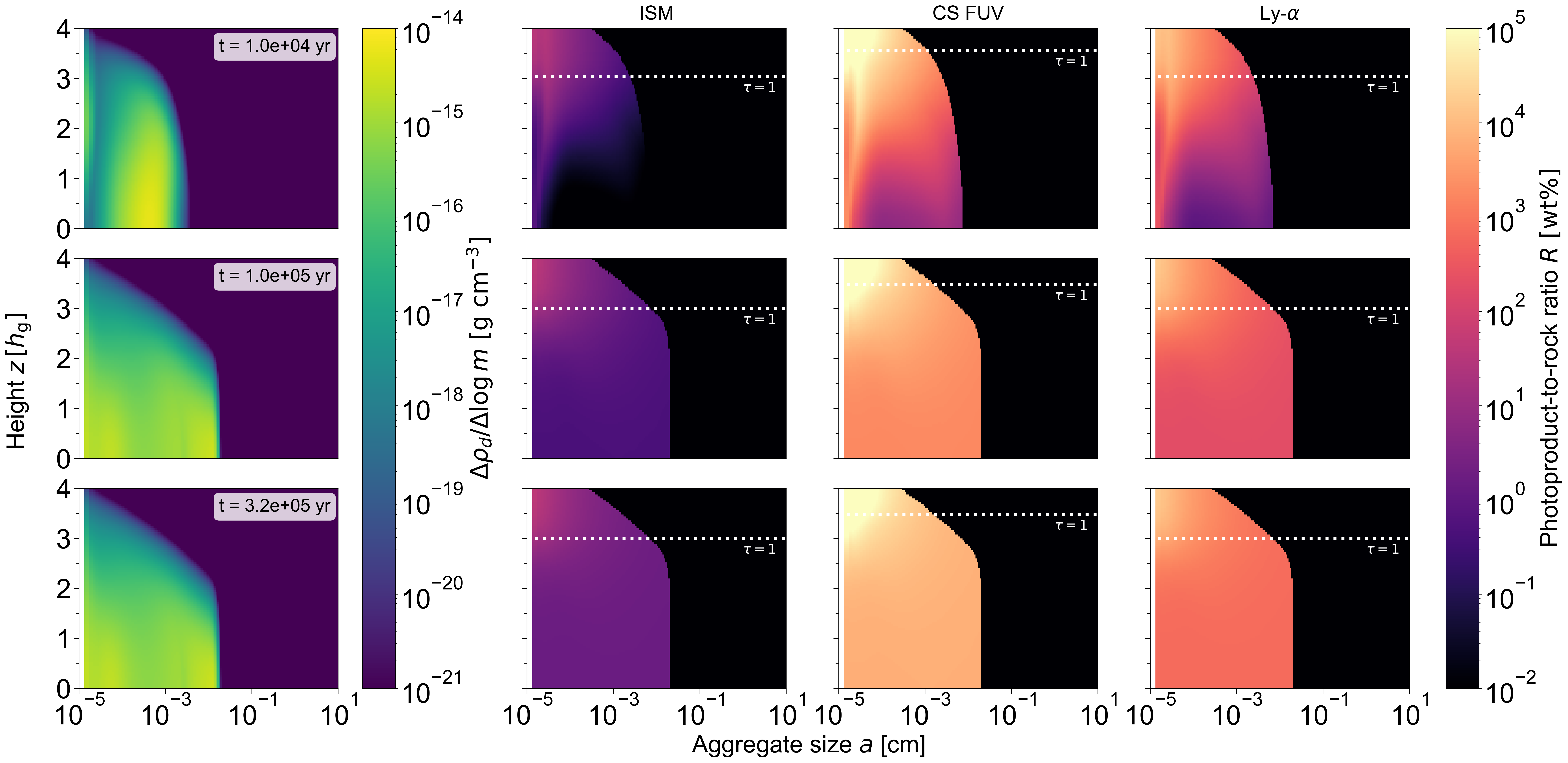}
  }
  \caption{Snapshots of the dust size distribution (first column) and photoproduct-to-rock ratio (second to fourth columns) for the \texttt{ism\_no\_lim}, \texttt{cs\_fuv\_no\_lim}, and \texttt{ly\_$\alpha$\_no\_lim} runs respectively, with all runs sharing the same dust size distribution. The top row shows the snapshot at $10^{4}$ yr, the middle at $10^{5}$ yr, and the bottom at $3.2 \times 10^{5}$ yr. The y-axis shows the height above the mid-plane (normalized to the gas scale-height $h_{\rm g}$), and the x-axis plots the aggregate size. The dotted lines in columns two and four show the location above the mid-plane where the vertical optical depth $\tau_{\perp} = 1$, whereas the dotted lines in column 3 show the location above the mid-plane where the line-of-sight optical depth $\tau_{\rm LOS} = 1$. Note the change in scale in the photoproduct-to-rock ratio color bar as compared to previous plots.}
  \label{fig results ISM, CS FUV, Ly-a combined no limit}
\end{figure*}

\section{Discussion} 
\label{sec: discussion}

\subsection{Validity of the optical depth model} 
\label{ssec: validity of the optical depth model}

Our treatment of dust fragmentation may overestimate the disk's opacity, as we assume that dust fragments from collisions are reduced to monomer sizes, whereas a realistic fragment size distribution may contain fewer small grains. This could allow UV radiation to penetrate deeper into the disk. However, the global photoproduct-to-rock ratio appears to be relatively insensitive to the precise optical depth of the disk. Comparing the $\texttt{ism\_fid}$, $\texttt{ism\_a4}$, and $\texttt{ism\_vf10}$ models in figure \ref{fig photoproduct-to-rock ratio time evolution all}, we see a difference of only a few percent in the final global photoproduct-to-rock ratio. Therefore, while introducing a size distribution for the fragments may alter the disk's opacity and FUV penetration depth, we do not expect it to significantly affect the global photoproduct production rate. 

More importantly, our calculation assumed that the UV photons are absorbed only by the dust particles in the disk. However, gas species can also contribute to the disk opacity as gas molecules are photodissociated and photoionized by UV photons \citep{van_dishoeck_photoprocesses_2006}. Here we consider the effect of gas absorption in the disk.

The disk's vertical optical depth due to gas absorption is given by $\tau_{\rm g} = \sum_{i} \sigma_{i} \int_z^\infty b_{i}\rho_{\rm g}/m_{i}\,dz'$, where $\sigma_i$, $b_i$, and $m_i$ are the absorption cross-section, the relative abundance with respect to the gas density, and the molecular mass of gas species $i$, respectively. The gas abundance of each species depends on the temperature and disk structure, as these determine the desorption of volatile ices from dust grains and the subsequent chemical evolution \citep{aikawa_evolution_1997, furuya_reprocessing_2014}. Since we do not include chemistry in our model, we instead estimate the gas abundances from the volatile ice inventory. Outside the snow line of a given species, the volatile is assumed to be frozen onto dust grains, whereas inside the snow line it sublimates into the gas phase. We therefore assume that the gas abundance interior to the snow line is set by the ice abundance exterior to the snow line.
Among the major volatile species in cometary ice are \ce{CO2}, \ce{CO}, and \ce{CH4}, with typical ice abundances relative to \ce{H2O} of $\sim10\,\%$ for \ce{CO2} and \ce{CO}, and $\sim1\,\%$ for \ce{CH4} \citep{mumma_chemical_2011}. At the Ly-$\alpha$ wavelength, however, their photodissociation cross-sections span $\sim10^{-20}$--$10^{-17}~\rm cm^2$ \citep{van_dishoeck_photoprocesses_2006}. Although \ce{CO2} is roughly an order of magnitude more abundant than \ce{CH4}, its absorption cross-section is several orders of magnitude smaller. Consequently, \ce{CH4} dominates the gas absorption of Ly-$\alpha$ photons. We therefore neglect the contributions from \ce{CO} and \ce{CO2} and approximate the gas opacity using only \ce{CH4}.

Using the adopted ice mass fraction of 0.5 and a comet-like ice composition \citep{mumma_chemical_2011}, in which \ce{CH4} comprises approximately $1\,\%$ of the ice mass, we obtain $b_{\ce{CH4}} = 0.01\times0.5\rho_{\rm d}/\rho_{\rm g}=5\times10^{-5}$. Taking $\sigma_{\ce{CH4}} = 1.8\times10^{-17}~\rm cm^2$ \citep{van_dishoeck_photoprocesses_2006} and $m_{\ce{CH4}} = 16m_{\rm H}\approx3.0\times10^{-23}~\rm g$, the gas optical depth becomes $\tau_{\rm g} \approx 30\,\mathrm{cm^2\,g^{-1}}\int_z^\infty \rho_{\rm g}\,dz' = 15\,\mathrm{cm^2\,g^{-1}}\Sigma_{\rm g}\operatorname{erfc}(z/(\sqrt{2}\,h_{\rm g}))$, where the complementary error function is $\operatorname{erfc}(x) = (2/\sqrt{\pi})\int_x^\infty e^{-t^2}\,dt$. Given the gas density and the height where $\tau_{\perp}=1$ ($z\approx3h_{\rm g}$) for the \texttt{ism\_fid} run, the gas optical depth is $\tau_{\rm g}\approx0.8$. While this is a rough estimate, we see that the gas optical depth is comparable to the dust optical depth for the fiducial case.

The above estimation suggests that organic matter can still be synthesized in strongly turbulent disks, though the rate of production might decrease by a factor of a few as the gas absorbs a part of the UV flux. In weakly turbulent disks, dust particles are less exposed to UV photons as the gas blocks the irradiation. This effect warrants a more detailed study. 

Finally, as the most abundant component of gas in a disk, it is worth noting the effect of molecular hydrogen gas absorption of FUV photons. Although it is possible for \ce{H2} gas to absorb FUV photons into the Lyman/Werner bands \citep{herczeg_origins_2006}, most of this pumping is limited to the wavelength range of $912$--$1110$ \AA$~$ \citep{draine_structure_1996}. Due to the exceedingly small photoabsorption cross sections of \ce{H2} molecules ($\sigma_{\rm abs,~H_{2}} \sim 10^{-24}$--$10^{-26} \rm cm^{2}$) \citep{van_dishoeck_photoprocesses_2006} to photons in the FUV range $1110$--$2070$ \AA$~$ \citep{hrodmarsson_photodissociation_2023}, the resulting optical depth due to molecular hydrogen gas for Ly-$\alpha$ photons calculated from our model is $\sim 3\times 10^{3}$ times smaller than the dust optical depth at the location where the latter reaches unity.

\subsection{Implications for organic synthesis and dust coagulation} 
\label{ssec: implications for organic synthesis and dust coagulation}

Photoproducts are converted into organic residue when the ice mantle sublimates as dust grains cross the water-ice snow line. While the composition of this residue that coats dust grains inside the snow line cannot be ascertained without a chemical model, ice photo-processing experiments by \cite{bernstein_organic_1995} revealed that the organic matter remaining after photolyzed ice analogs were warmed to room temperature contained hexamethylenetetramine (HMT), ethers and alcohols, and ketones and amides. These three groups made up about 60\%, 20\%, and 20\% of the total organic residue at 300 K, respectively. \cite{munoz_caro_uv-photoprocessing_2003} also confirmed the presence of HMT as the most abundant component of organic residue when \ce{CH3OH}-containing ice was photo-processed. They found that HMT is formed at room temperature, and also detected carboxylic acid salts, amides, and esters in the residue. Despite several experimental studies investigating COM formation in astrophysical conditions, reaction networks forming COMs in such ice analogs are not well understood. Photo-processing of ice can result in several more proposed products, such as sugars, lipids, amino acids, amines, and heterocycles \citep{ciesla_organic_2012, munoz_caro_photochemistry_2025}.

While the exact composition of the organic mantle that would result from the photoproducts in our simulation is uncertain, experiments show that the residue is not fully volatile, and survives at temperatures up to 300 K \citep{bernstein_organic_1995}. This stability is supported by matrix effects and H-bonding in the residue, enabling species to sublimate at higher temperatures than their pure counterparts \citep{munoz_caro_uv-photoprocessing_2003}. Therefore, even at distances close to the central star, grains maintain their organic mantles. 

Assuming that these organic mantles increase the sticking efficiency of dust, as proposed by \cite{homma_rocky_2019}, they could enable more efficient dust growth in the inner disk. \cite{homma_rocky_2019} found that organic-mantled grains with an organic mass fraction of $\approx 5\%$ may grow beyond the fragmentation barrier through collisions. This corresponds to an organic-to-rock ratio of $\approx 5\%$, which we therefore take as the threshold required for organics-assisted growth. As described in section \ref{sec: intro}, we define photoproducts as the fraction of irradiated molecules that ultimately form refractory organic matter following radial drift and warming of the ices. The production efficiency $\epsilon$ in equation \eqref{eq source term for organic matter} accounts for the number of organic molecules produced per incident photon, and is therefore already incorporated into the calculated photoproduct mass. Therefore, the photoproduct-to-rock ratio can be directly compared with the required organic-to-rock ratio of $\approx 5\%$ without applying an additional conversion efficiency.

Comparing the global photoproduct-to-rock ratio for each radiation source (figure \ref{fig global time evolution of photoproduct-to-rock ratio}), we see that the \texttt{ism\_fid} and \texttt{cs\_fuv\_fid} models produce insufficient photoproducts to reach the required organic-to-rock ratio of 5\% after radial drift. However, taking into account the Ly-$\alpha$ emission from the central star, we see that the \texttt{ly\_$\alpha$\_fid} model exceeds the required organic ratio. Therefore, our results reveal that dust grains irradiated by the central star will have sufficiently thick organic mantles once drifted into the inner disk, leading to more efficient dust coagulation.

However, it is important to note that as grains drift into the inner region of the disk, they are subject to a variety of processes that can have destructive effects on their organic mantles. Unlike in the outer disk, where radiation serves only to increase the photoproduct abundance, interior to the snow line, radiation also has a destructive effect on synthesized organics. Dust drifting radially inward and passing the snow line is exposed to an increasingly intense radiation field from the central star, which can destroy or alter the organic mantle through photolysis. \cite{siebenmorgen_destruction_2010} investigated the destruction of polycyclic aromatic hydrocarbons (PAHs) in disks around T Tauri stars, and found that destruction via FUV radiation is efficient inside 1 au. More energetic photons (extreme ultraviolet and X-ray photons) can also destroy PAHs at larger radii. However, they proposed that vertical mixing of dust grains could replenish PAHs to the surface layer, where most of the destruction takes place. Similarly, \cite{anderson_destruction_2017} found that carbon depletion in the mid-plane is less effective than at the surface, and depends upon dust transport processes within the disk. This is consistent with \cite{klarmann_radial_2018}, who studied the effect of dust transport on carbon depletion in disks, and found that carbon depletion cannot be entirely attributed to mechanisms such as photolysis that occur mainly at the disk surface. \cite{vaikundaraman_refractory_2025} further demonstrated the importance of vertical turbulent mixing in determining the carbon depletion timescale, as efficient vertical diffusion transports carbon from the mid-plane to the surface layers of the disk, where they are more readily destroyed. 

In addition, other processes such as oxidation \citep{lee_solar_2010} and thermal decomposition \citep{nakano_evaporation_2003, houge_burned_2025} can also play a role in the destruction of organic molecules, especially in the inner disk. The oxidation of carbonaceous material via O-bearing species and radicals can erode carbon off dust grains. However, \cite{anderson_destruction_2017} notes that this effect is modest, and only operates near the surface of the disk. Moreover, oxidation requires high ($\gtrsim 1200$ K) temperatures due their large activation energies, and is therefore efficient only in the very inner disk \citep{gail_radial_2002}. 

\cite{kress_sootline_2010} proposed the idea of a 'soot line' as the border inside which organic matter is irreversibly destroyed via thermal destruction of molecules into smaller volatiles and their subsequent sublimation. In the early phase of star formation, accretion heating may have placed this soot line at tens of au at the disk mid-plane \citep{li_earth_2021}. However, as the disk evolves and stellar irradiation becomes the dominant heating source, the soot line migrates inward, interior to 1 au \citep{binkert_carbon_2023}. The thermal decomposition of refractory organics was studied by \cite{houge_burned_2025}, who found that refractory organic matter decomposes at temperatures around 300--500 K. However, different types of organic matter are destroyed at different temperatures \citep{gail_spatial_2017, okamoto_effects_2024}, therefore the location of this thermal-processing boundary depends on the type of carbonaceous material considered. A soot line of 450 K, the temperature required for the out-gassing of refractory organic material given by \cite{gail_spatial_2017}, is located at a distance of 0.38 au for our temperature model. The soot line is therefore inside the water snow line, and we expect that the organic mantles formed on dust grains after radial drift and ice sublimation survive long enough to aid dust growth in the region between the snow line and the soot line. 

In conclusion, while the very inner regions of the disk may be hostile to organic material, we expect that these destructive processes do not entirely destroy the synthesized organic mantles. Moreover, photolysis and oxidation are shown to be most efficient at the disk surface, and therefore destruction via these mechanisms depend on the efficiency of turbulence in the disk. Dust near the mid-plane are more likely to maintain their organic mantles in disks with inefficient vertical transport. In order to more thoroughly investigate the evolution of dust inside the snow line, it would be important to introduce a chemical model including synthesis and destruction via the above processes to illuminate how organic material influences dust growth in the inner disk. We leave this to future work. 

{Additionally, we briefly discuss the validity of neglecting radial drift in our model. In order to assess the effect of radial drift, we compare the radial drift timescale, $t_{\rm drift}$, with the simulation timescale. The radial drift timescale is given as $t_{\rm drift} = r/v_r$. For the \texttt{ly\_$\alpha$\_fid} run, the largest grains ($a_{\rm max}=0.02~\rm cm$) are roughly estimated to have a drift timescale of $t_{\rm drift}\approx 4\times10^{5}$ yr, while the simulation runs for $t=10^{6}$ yr. This drift timescale represents the shortest drift timescale among the grains considered, as smaller grains drift more slowly. Furthermore, at $t=t_{\rm drift}=4\times10^{5}$ yr, the \texttt{ly\_$\alpha$\_fid} run already reaches a global photoproduct-to-rock ratio of $6\%$, exceeding the required threshold of $\approx 5\%$ for aggregates to become sufficiently sticky. Thus, even accounting for radial drift of the largest grains, our main conclusion regarding organic enrichment and the conditions required for aggregate growth remains unchanged.}


\section{Conclusions}\label{sec: conclusions}

In this work we have presented results of photoproduct formation via three different sources of UV radiation on dust aggregates in protoplanetary disks. We simulated the evolution of the vertical dust size distribution, including the effects of collisional growth, fragmentation, and diffusion. We simulated three different UV sources, namely, the interstellar medium, the disk's central star, and nearby massive stars, in order to quantify the amount of photoproducts produced on the grains in each case. We compared our results with the organic-to-rock ratio of $\approx 5\%$ required for enhanced grain sticking proposed by \cite{homma_rocky_2019}, after grains drift inward. Our main findings are as follows:

\begin{enumerate}
    \item While the ISM is the most commonly assumed radiation source for protoplanetary disks, two other sources, the central star and nearby massive stars, are significant sources of ionizing radiation that should not be ignored, especially in the context of photo-production of radicals, ions, and intermediary molecules for organic synthesis (subsection \ref{ssec: UV radiation}). {Ly-$\alpha$ radiation from the central star and FUV radiation from nearby massive stars produce photoproduct-to-rock ratios of up to 10\%.}
    \item The distinction between Ly-$\alpha$ and FUV continuum photons emitted by the central star is an important one. As Ly-$\alpha$ photons interact with both chemical species and the disk layers differently than FUV continuum photons, their contribution to photoproduct formation should not be understated (subsection \ref{ssec: fiducial models}). Moreover, due to additional processing by the hydrogen layer, Ly-$\alpha$ photons are able to penetrate the disk in much the same way as radiation from the ISM and nearby massive stars (subsubsection \ref{sssec: central star Ly-alpha}). 
    \item Although dust aggregates near the mid-plane are not directly exposed to the radiation incident on the disk, they experience a cycle of growth, fragmentation, and turbulent diffusion that allows them to form photoproducts in their mantles. As a result, irradiated grains are found throughout the vertical extent of the disk (figure \ref{fig results ISM, CS FUV, Ly-a, NMS combined fiducial}).
    \item Ly-$\alpha$ radiation provides grains with enough photoproducts to {reach an organic-to-rock ratio of nearly 10\% after radial drift, substantially exceeding the $\approx 5\%$ threshold. The thick organic mantles resulting from this high organic abundance may enhance grain stickiness,} leading to more efficient planet formation in the inner disk (figure \ref{fig global time evolution of photoproduct-to-rock ratio}, subsection \ref{ssec: implications for organic synthesis and dust coagulation}).
\end{enumerate}

Our results show that Ly-$\alpha$ radiation from the central star sufficiently irradiates dust grains to produce photoproducts for organic synthesis. This has certain implications for planetesimal formation. As highly irradiated dust grains drift inward to the warmer parts of the disk (inside the snow line), the warming of their ice mantle sparks the production of organic matter on their surface. Assuming these mantles are sticky, the dust grains in the inner disk, aided by their organic mantles, coagulate and grow more efficiently via collisions, and are potentially able to overcome barriers to dust growth such as fragmentation. 

However, as we do not include a chemical model in our calculation, it is unclear what kinds of organic matter are formed and how they are destroyed. In the future, to investigate the influence of organic mantles on dust growth, we hope to simultaneously model the radial drift of dust grains and the chemical evolution of their organic mantles. This will allow us to investigate the types and species of organic matter formed, while also accounting for their destruction in the inner disk, thereby providing a more complete treatment of the chemical and dynamical evolution of dust.


\section*{Funding}
This work was supported by JSPS KAKENHI Grant numbers JP23K25923, JP23H00143, and JP25K00025.

\begin{ack}
We thank Uma Gorti, Kenji Furuya, Yoko Kebukawa, and Yasuhiro Oba for useful discussions that helped improve this work.
\end{ack}

{
\appendix 

\section{Supplementary figures}
\label{sec:appendix}

This appendix presents the results for the \texttt{ly\_$\alpha$\_a4} and \texttt{ly\_$\alpha$\_vf10} runs, shown in figures \ref{fig results Ly-a alpha4} and \ref{fig results Ly-a vfrag10} respectively. Additionally, the results for the \texttt{ism\_a4\_no\_lim} and \texttt{ly\_$\alpha$\_a4\_no\_lim} runs are shown in figure \ref{fig results ISM, Ly-a combined alpha4 no limit}, while the results of the \texttt{ism\_vf10\_no\_lim} and \texttt{ly\_$\alpha$\_vf10\_no\_lim} runs are shown in figure \ref{fig results ISM, Ly-a combined vfrag10 no limit}.

\begin{figure}
  {\centering
  \includegraphics[width=\linewidth, bb = 0 0 2307.6 1833.12]{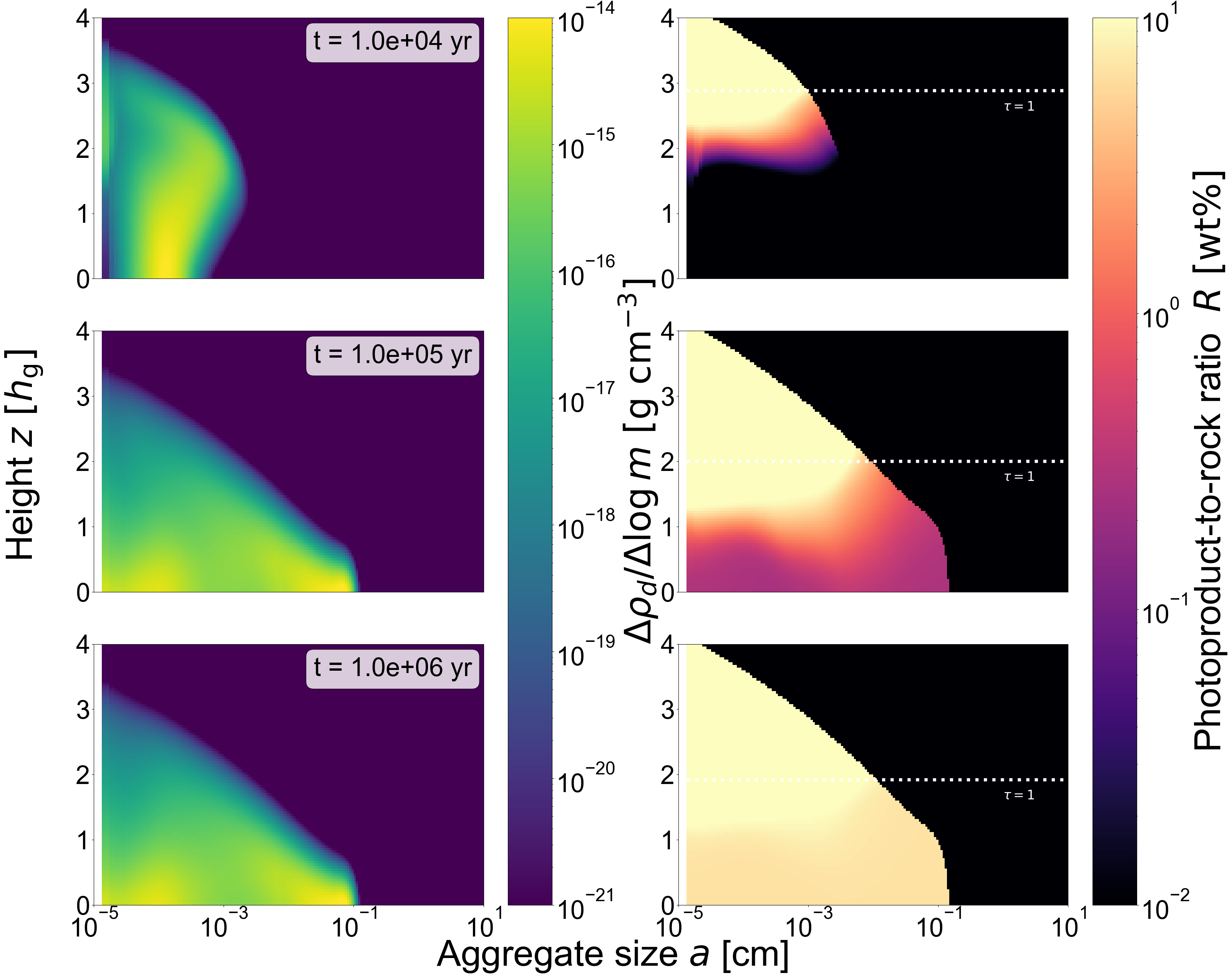}
  }
  \caption{Same as figure \ref{fig results ISM alpha4} but for the \texttt{ly\_$\alpha$\_a4} run.}
  \label{fig results Ly-a alpha4}
\end{figure}

\begin{figure}
  {\centering
  \includegraphics[width=\linewidth, bb = 0 0 2307.6 1833.12]{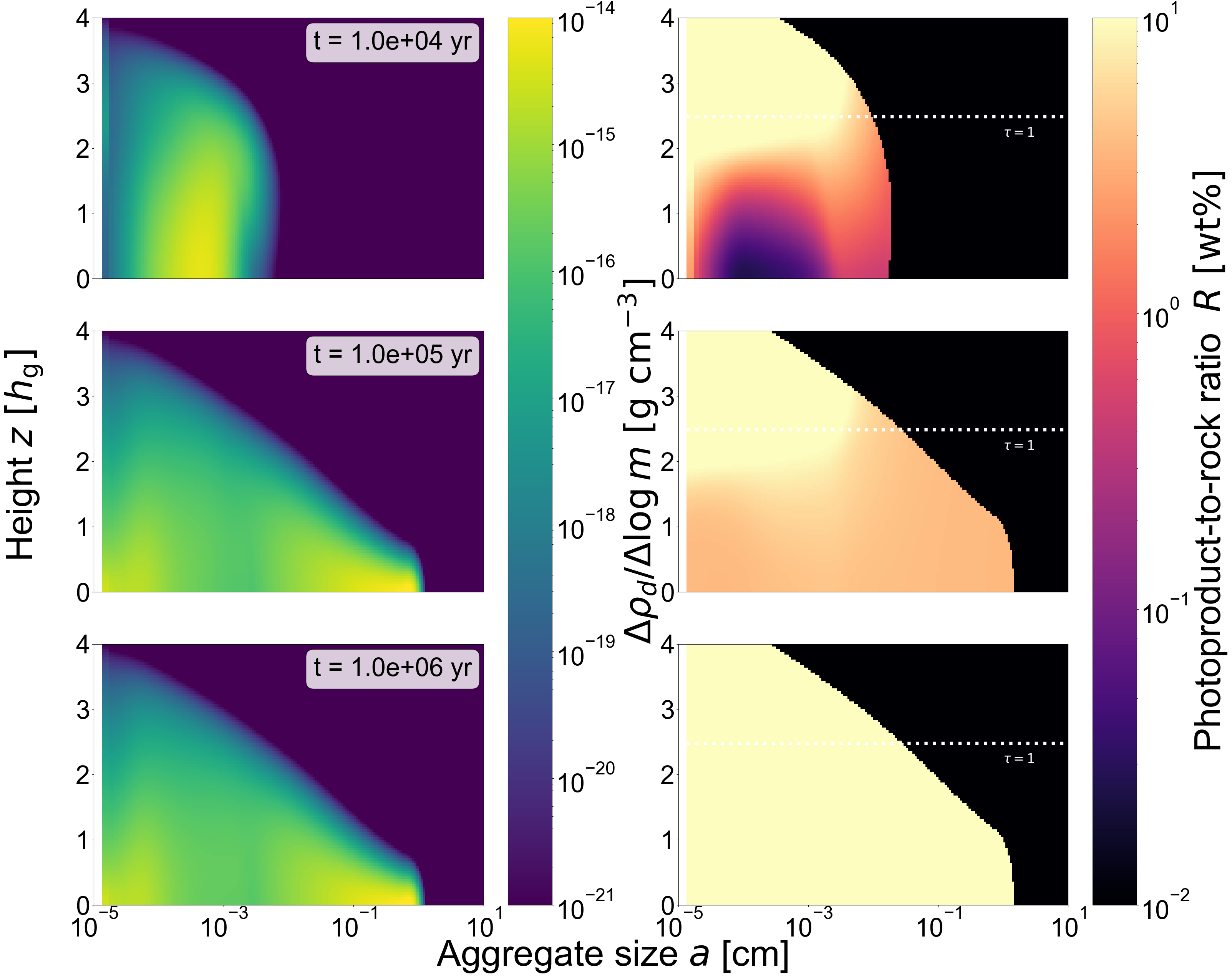}
  }
  \caption{Same as figure \ref{fig results ISM alpha4} but for the \texttt{ly\_$\alpha$\_vf10} run.}
  \label{fig results Ly-a vfrag10}
\end{figure}

\begin{figure*}[htbp]
  \begin{center}
  \includegraphics[width=0.7\textwidth, bb = 0 0 1568.16 1016.16]{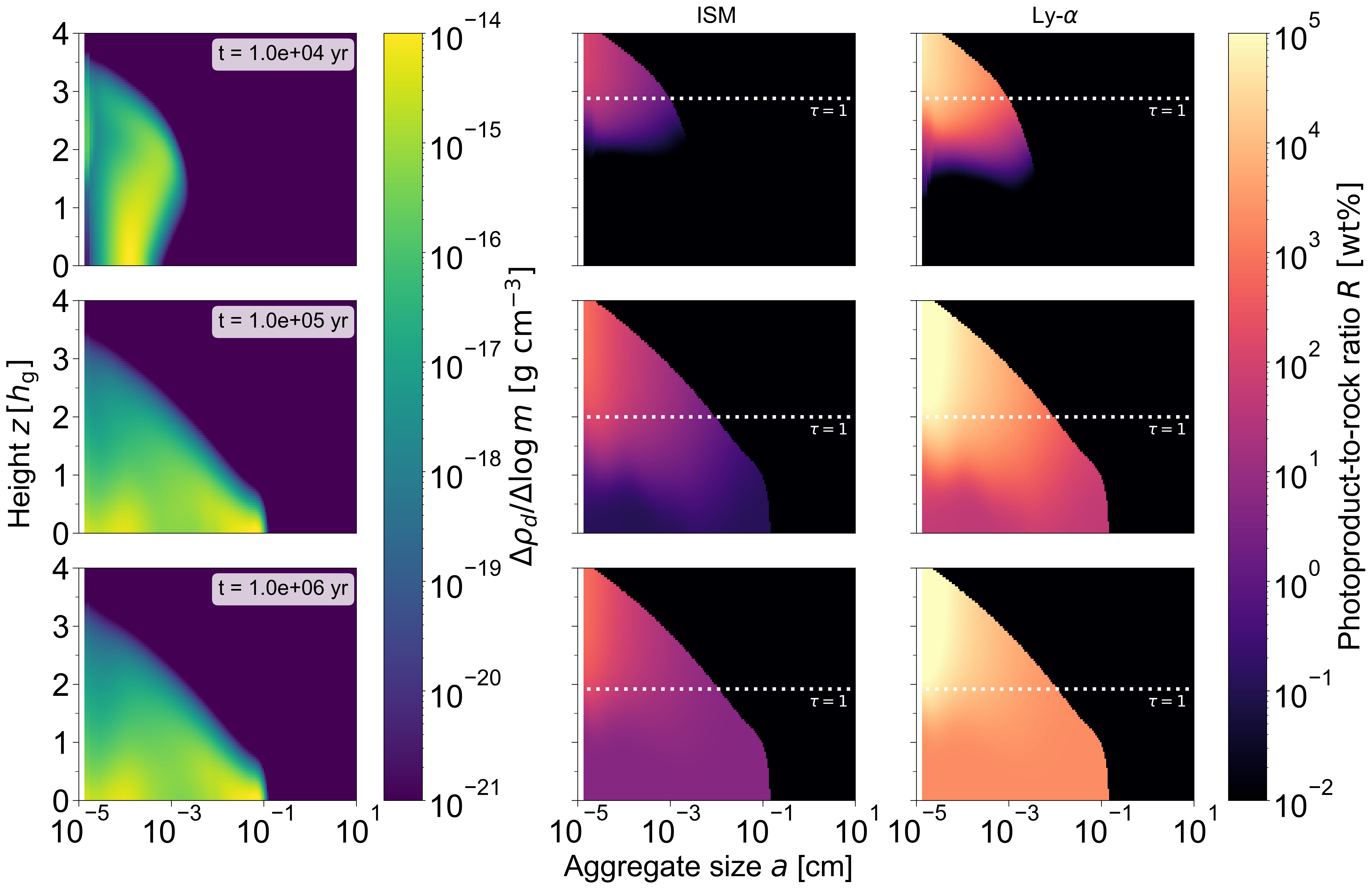}
  \end{center}
  \caption{Snapshots of the dust size distribution (first column) and photoproduct-to-rock ratio (second and third columns) for the \texttt{ism\_a4\_no\_lim} and \texttt{ly\_$\alpha$\_a4\_no\_lim} runs respectively, with both runs sharing the same dust size distribution. The top row shows the snapshot at $10^{4}$ yr, the middle at $10^{5}$ yr, and the bottom at $10^{6}$ yr. The y-axis shows the height above the mid-plane (normalized to the gas scale-height $h_{\rm g}$), and the x-axis plots the aggregate size. The dotted lines in columns two and three show the location above the mid-plane where the vertical optical depth $\tau_{\perp} = 1$. Note the change in scale in the photoproduct-to-rock ratio color bar as compared to previous plots.}
  \label{fig results ISM, Ly-a combined alpha4 no limit}
\end{figure*}

\begin{figure*}[htbp]
  \begin{center}
  \includegraphics[width=0.7\textwidth, bb = 0 0 1568.16 1016.16]{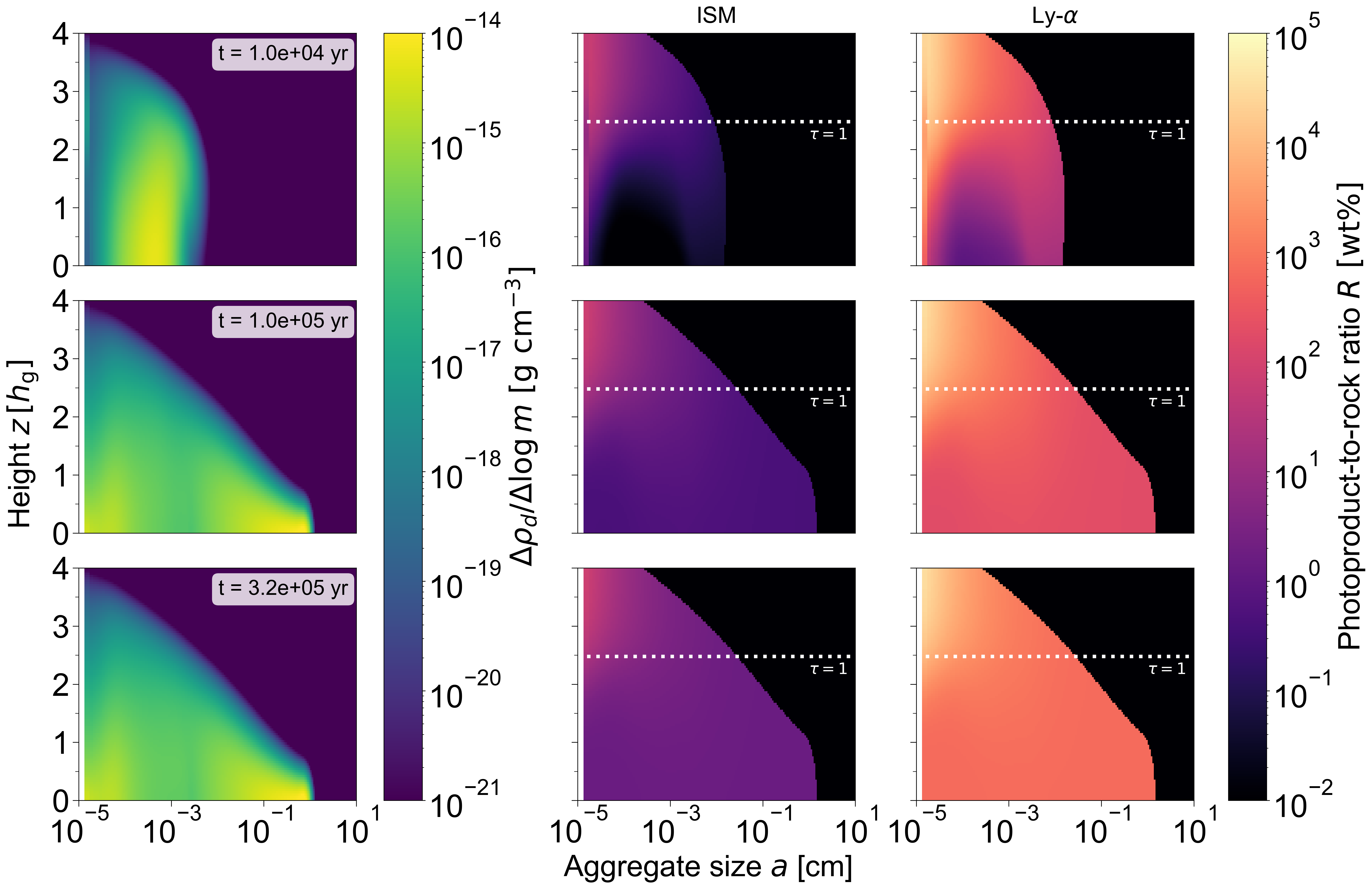}
  \end{center}
  \caption{Snapshots of the dust size distribution (first column) and photoproduct-to-rock ratio (second and third columns) for the \texttt{ism\_vf10\_no\_lim} and \texttt{ly\_$\alpha$\_vf10\_no\_lim} runs respectively, with both runs sharing the same dust size distribution. The top row shows the snapshot at $10^{4}$ yr, the middle at $10^{5}$ yr, and the bottom at $3.2 \times 10^{5}$ yr. The y-axis shows the height above the mid-plane (normalized to the gas scale-height $h_{\rm g}$), and the x-axis plots the aggregate size. The dotted lines in columns two and three show the location above the mid-plane where the vertical optical depth $\tau_{\perp} = 1$. Note the change in scale in the photoproduct-to-rock ratio color bar as compared to previous plots.}
  \label{fig results ISM, Ly-a combined vfrag10 no limit}
\end{figure*}
}

\bibliographystyle{apj}
\bibliography{bibliography}

\end{document}